\documentclass[sigconf]{acmart}

\renewcommand\footnotetextcopyrightpermission[1]{} % removes footnote with conference information in first column
\usepackage{graphicx}
\DeclareGraphicsExtensions{.pdf,.png,.jpg,.jpeg}

\graphicspath{{figures/}{pictures/}{images/}{./}} % where to search for the images

\usepackage{enumitem}
\usepackage{xcolor}
\usepackage{subcaption}
\usepackage{tabularx}
\usepackage{arydshln}
\usepackage{placeins}

\newcommand{\supplemental}[1]{Supplemental {#1}}
\newcommand{\ngram}[1]{{\fontsize{8.5}{8}\texttt{#1}}}
\newcommand{\question}[1]{
  \vspace{0.25em}
  \noindent {\bf \em #1}
}

\begin{document}

\title{Analyzing Who and What Appears in a Decade of \\ US Cable TV News}
\author{James Hong, Will Crichton, Haotian Zhang, Daniel Y. Fu, Jacob Ritchie}
\author{Jeremy Barenholtz, Ben Hannel, Xinwei Yao, Michaela Murray}
\author{Geraldine Moriba, Maneesh Agrawala, Kayvon Fatahalian}
\affiliation{
  \institution{Stanford University}
}

\renewcommand{\shortauthors}{Hong, et al.}

\begin{abstract}

Cable TV news reaches millions of U.S. households each day, meaning that
decisions about who appears on the news and what stories get covered can
profoundly influence public opinion and discourse.  We analyze a data set of
nearly 24/7 video, audio, and text captions from three U.S. cable TV networks
(CNN, FOX, and MSNBC) from January 2010 to July 2019. Using machine learning
tools, we detect faces in 244,038 hours of video, label each face's presented
gender, identify prominent public figures, and align text captions to audio. We
use these labels to perform screen time and word frequency analyses. For
example, we find that overall, much more screen time is given to male-presenting
individuals than to female-presenting individuals (2.4x in 2010 and 1.9x in
2019). We present an interactive web-based tool, accessible at 
\url{https://tvnews.stanford.edu}, that allows the general public
to perform their own analyses on the full cable TV news data set.

\end{abstract}




\begin{teaserfigure}
  \centering
  \vspace{-0.05in}
  \includegraphics[width=\linewidth]{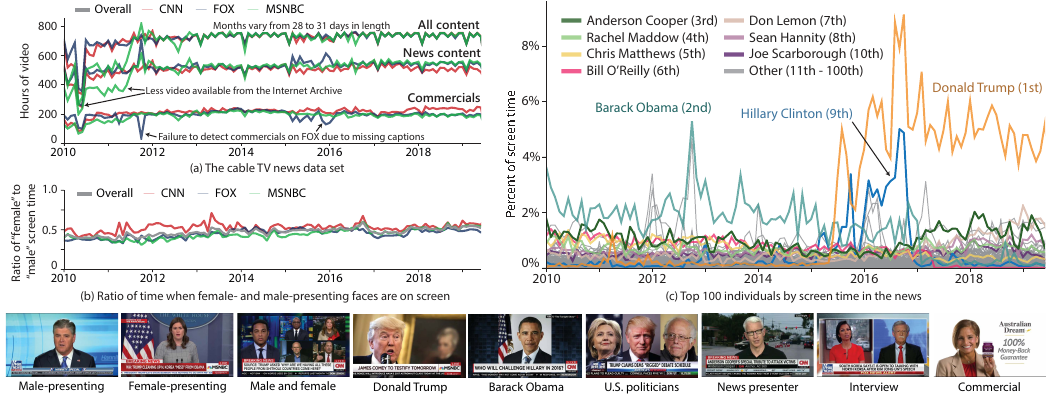}
  \vspace{-0.05in}
  \caption{
    \textbf{(a)} Our data set contains over 244,000 hours of video aired on CNN,
    FOX, and MSNBC from January 1, 2010 to July 23, 2019. The screen time of news
    content (and commercials) in our data set is stable from 2012 onwards,
    representing near 24/7 coverage.
    \textbf{(b)} The ratio of time of when female-presenting faces are on screen
    to when male-presenting faces are on screen is 0.48 to 1 on average,
    but has risen from 0.41 (to 1) to 0.54 (to 1) over the decade.
    \textbf{(c)} The top 100 people by face screen time in the data set, with names of the top 10 given.
    Of the top 100 people, 18 are U.S. politicians and 85 are news presenters (3 are both).
  }
  \label{fig:teaser}
\end{teaserfigure}

\maketitle
\pagestyle{plain}

\section{Introduction}

Cable TV news reaches millions of U.S. households each day, and
profoundly influences public opinion and discourse on current
events~\cite{tvnewsPew}.  While cable TV news has been on air for over
40 years, there has been little longitudinal analysis of its visual
aspects. As a result, we have little understanding of {\em who}
appears on cable TV news and {\em what} these individuals talk about.

Consider questions like, {\em
  What is the screen time of men vs. women? Which political candidates
  and news presenters receive the most screen time? How are victims
  and perpetrators of violence portrayed? Which foreign
  countries are discussed the most? Who is on screen when different
  topics are discussed?}

In this paper, we demonstrate that it is possible to answer such questions by
analyzing a data set comprised of nearly 24/7 coverage of video, audio, and text
captions from three major U.S. cable TV news channels -- CNN, FOX (News) and
MSNBC -- over the last decade (January 1, 2010 to the present). The data set was
collected by the Internet Archive's TV News Archive~\cite{tvnewsarchive}. We
focus our analysis (and validation) between January 2010 to July 2019, which
includes 244,038 hours (equivalent to about 27.8 years) of footage. Using
programmatic and machine learning tools, we label the data set -- e.g., we
detect faces, label their presented gender, identify prominent public figures,
align text captions to audio, and detect commercials. We scope our findings to
the \emph{news programming} portion of the data set (\autoref{fig:teaser}a),
accounting for 72.1\% of the video (175,858 hours) compared to 27.9\% for
commercials (68,179 hours).

Each of the resulting labels has a temporal extent, and we use these extents to
compute the {\em screen time} of faces and to identify when individuals are on
screen and when words are said. We show that by analyzing the screen time of
faces, counting words in captions, and presenting results in the form of
time-series plots, we can reveal a variety of insights, patterns, and trends
about the data. To this end, we adopt an approach similar to the Google N-gram
viewer~\cite{ngrams}, which demonstrated the usefulness of word frequency
analysis of 5.2 million books and print media from 1800 to 2000 to many
disciplines, as well as to the GDELT AI Television
Explorer~\cite{gdelt_ai_explorer}, which enables analysis of cable TV news
captions and on screen objects (but not people). The goal of our work is to
enable similar analyses of cable TV news video using labels that aid
understanding of who is on screen and what is in the captions.

Our work makes two main contributions.
\begin{itemize}[noitemsep,topsep=0pt]

  \item We demonstrate that analyzing a decade (January 1, 2010 to July 23,
  2019) of cable TV news video generates a variety of insights on a range of
  socially relevant issues, including gender balance (\autoref{sec:face}),
  visual bias (\autoref{sec:portrayal}), topic coverage (\autoref{sec:text}),
  and news presentation (\autoref{sec:faces_and_text}). The details of our data
  processing, labeling pipeline, and validation for these analyses are described
  in~\supplemental{1}.

  \item We present an interactive, web-based data analysis interface
  (\autoref{sec:tool}; available at \url{https://tvnews.stanford.edu}), that
  allows users to easily formulate their own their analysis queries on our
  annotated data set of cable TV news (\autoref{sec:discussion}). Our analysis
  interface updates daily with new cable TV news video and allows the general
  public and journalists to perform their own exploratory analyses on the full
  the TV news data set. Our data processing code is available as open
  source.

\end{itemize}

\begin{figure*}[!tbp]
  \centering
  \includegraphics[width=\textwidth]{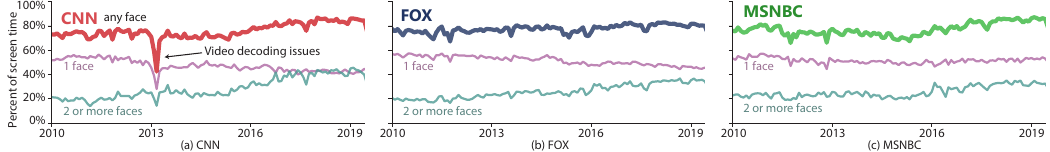}
  \caption{The percentage of time when at least one face appears on screen has increased on all three channels over the decade (thick lines), with most of the increase occurring between 2015 and 2018. The amount of time when multiple faces are on screen has also increased on all three channels, however the percentage of time with only one face on screen has declined on CNN and FOX, and stagnated on MSNBC.}
  \label{fig:num_faces_onscreen}
\end{figure*}

\section{Who is in the news?}
\label{sec:face}

People are an integral part of the news stories that are covered, how they are
told, and who tells them. We analyze the screen time and demographics of faces
in U.S. cable TV news.

\question{How much time is there at least one face on screen?} We detect faces
using the MTCNN~\cite{mtcnn} face detector on frames sampled every three seconds
(\supplemental{1.3}). Face detections span a wide range of visual contexts
ranging from in-studio presenters/guests, people in B-roll footage, or static
infographics. Overall, we detect 263M total faces, and at least one face appears
on screen 75.3\% of the time. Over the decade, the percentage of time with at
least one face on screen has risen steadily from 72.9\% in 2010 to 81.5\% in
2019 and is similar across all three
channels~(\autoref{fig:num_faces_onscreen}).

In the same time period, we also observe an increase in the average number of
faces on screen. On CNN and FOX the amount of time when only one face is on
screen has declined, while it has remained constant on MSNBC. On all three
channels, the amount of time when multiple faces (2 or more) are on screen
simultaneously has risen. This accounts for the overall increase in time when at
least one face is on screen, though we do not analyze which types of content
(with no faces on screen) that this footage is replacing. We note that while the
average number of faces has increased in news content, the average number of
faces on screen in commercials has remained flat since 2013
(\supplemental{2.1.1}).

\begin{figure}[!tbp]
  \centering
  \includegraphics[width=\columnwidth]{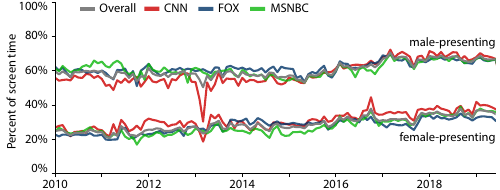}
  \caption{The percentages of time when male-presenting and female-presenting faces are
  on screen are similar on all three channels, and have increased over the decade with the rise in all faces noted
  in~\autoref{fig:num_faces_onscreen}. Because male- and female-presenting faces can be on
  screen simultaneously, the lines can add to over 100\%.}
  \label{fig:gender_time}
\end{figure}

\question{How does screen time of male-presenting individuals compare to
female-presenting individuals?} We estimate the presented binary gender of each
detected face using a nearest neighbor classifier trained on
FaceNet~\cite{facenet} descriptors (\supplemental{1.4}). Overall,
female-presenting faces are on screen 28.7\% of the time, while male-presenting
faces are on screen 60.2\% of the time, a 0.48 to 1 ratio
(\autoref{fig:gender_time}). These percentages are similar across channels and
have slowly increased for both groups (similar to how the percentage of time any
face is on screen has increased). The ratio of female- to male-presenting screen
time has increased from 0.41 (to 1) to 0.54 (to 1) over the decade
(\autoref{fig:teaser}b). While the upward trend indicates movement towards
gender parity, the rate of change is slow, and these results also reinforce
prior observations on the under-representation of women in both
film~\cite{geenadavis} and news media~\cite{gmmp}.

We acknowledge that our simplification of presented gender to a binary quantity
fails to represent transgender or gender-nonconforming
individuals~\cite{keyes2018,hamidi2018}.  Furthermore, an individual's presented
gender may differ from their actual gender identification. Despite these
simplifications, we believe that automatically estimating binary presented
gender labels is useful to improving understanding of trends in gender
representation in cable TV news media.

\begin{figure}[!tbp]
  \centering
  \includegraphics[width=\columnwidth]{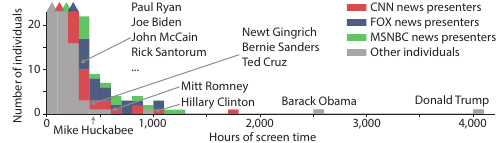}
  \caption{Distribution of individuals' screen time, separated by presenters on
  each channel and non-presenters (stacked). 65\% of individuals with 100+ hours
  of screen time are news presenters. The leading non-presenters are annotated --- see \autoref{fig:top_hosts} for the top news presenters.
  Note: the three leftmost bars are truncated and the truncated portion includes presenters from all three channels.}
  \label{fig:person_histogram}
\end{figure}

\begin{figure*}[!tbp]
  \centering
  \includegraphics[width=\textwidth]{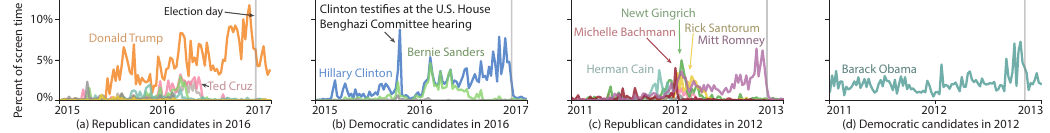}
  \caption{Screen time of U.S. presidential candidates during the campaign and primary season of the 2016 and 2012 elections. \textbf{(a)} In 2016, Donald Trump received significantly more screen time than the other Republican candidates. \textbf{(b)} Hillary Clinton and Bernie Sanders received nearly equal screen time during the competitive primary season (January-May 2016). \textbf{(c)} In 2012, Mitt Romney did not decisively overtake the other Republican candidates in screen time until he became the presumptive Republican nominee.}
  \label{fig:primaries}
\end{figure*}

\question{Which public figures receive the most screen time?} We estimate the
identity of faces detected in our data set using the Amazon Rekognition
Celebrity Recognition API~\cite{amazonrekognition}. For individuals who are not
currently included (or not accurately detected) by the API, we train our own
classifiers using FaceNet~\cite{facenet} descriptors. See \supplemental{1.5} for
details.

We identify 1,260 unique individuals who receive at least 10 hours of screen
time in our data set. These individuals account for 47\% of the 263M faces that
we detect in the news content and are on screen for 45\% of screen time. The top
individual is Donald Trump, who rises to prominence in the 2016 presidential
campaigning season and his presidency (\autoref{fig:teaser}c). Barack Obama is
second, with 0.63$\times$ Trump's screen time, and is prevalent between 2010
(the start of the data set) and 2017 (the end of his second term). Besides U.S.
presidents, the list of top individuals is dominated by politicians and news
presenters (e.g., anchors, daytime hosts, field reporters, etc.)
(\autoref{fig:person_histogram}).

\question{How much screen time do political candidates get before an election?}
During the 2016 Republican presidential primaries, Donald Trump consistently
received more screen time than any other candidate (\autoref{fig:primaries}a).
In the competitive months of the primary season, from January to May 2016, Trump
received 342 hours of screen time, while his closest Republican rival, Ted Cruz,
received only 130 hours. In the same timespan, the leading Democratic
candidates, Hillary Clinton and Bernie Sanders received more equal screen time
(164 hours compared to 139 hours for Clinton); both received far more screen
time than the other Democratic primary candidates (\autoref{fig:primaries}b).
Comparing the two presidential nominees during the period from January 1, 2016
to election day, Trump received 1.9$\times$ more screen time than Clinton.

Unlike Trump in 2016, in the run up to the 2012 presidential election, Mitt
Romney (the eventual Republican nominee) did not receive as dominating an amount
of screen time (\autoref{fig:primaries}c). Other Republican candidates such as
Herman Cain, Michelle Bachmann, Newt Gingrich, and Rick Santorum have higher
peaks than Romney at varying stages of the primary season, and it is not until
April 2012 (when his last rival withdraws) that Romney's screen time decisively
overtakes that of his rivals. For reference,~\autoref{fig:primaries}d shows
Barack Obama's screen time during the same period. As the incumbent president up
for re-election, Obama had no significant primary challenger. Obama received
more screen time throughout 2011 than Romney because, as the president, he is in
the news for events and policy actions related to his duties as president (e.g.,
U.S. missile strikes in Libya, job growth plan, etc.). In 2012, however, Obama
and Romney are comparable. The overall trends are similar when viewed by
channel, with Trump dominating screen time in 2016 on all three channels
(\supplemental{2.1.3}).

\begin{figure*}[!tbp]
  \centering
  \includegraphics[width=\textwidth]{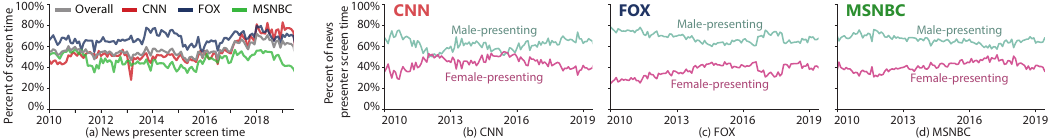}
  \caption{
  \textbf{(a)} The percentage of time when a news presenter is on screen has remained mostly flat on FOX and MSNBC, but has risen by 13\% on CNN since 2016.
  \textbf{(b-d)} Within each channel, the screen time of news presenters by presented-gender (as a percentage of total news presenter screen time) varies across the decade. CNN reaches parity in January-June 2012 and May-August 2015, but has since diverged. Because male- and female-presenting news presenters can be on screen simultaneously, the lines can add to over 100\%.}
  \label{fig:host_gender_time}
\end{figure*}

\begin{figure*}[!tbp]
  \centering
  \includegraphics[width=\textwidth]{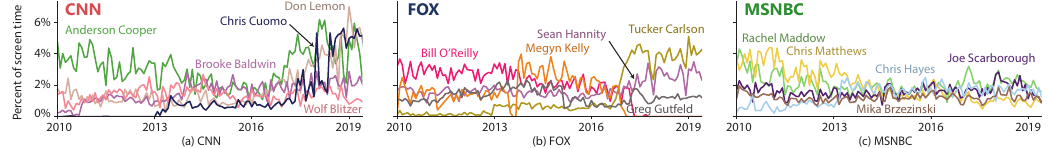}
  \caption{Screen time of the top five presenters on each channel. Since 2016, several of the top presenters on CNN have dramatically risen in screen time. Following Bill O'Reilly's firing and Megyn Kelly's departure from FOX in 2017, Sean Hannity and Tucker Carlson have risen in screen time. Since 2013, the variation in screen time among the top five hosts on MSNBC has been low compared to CNN and FOX.}
  \label{fig:top_hosts}
\end{figure*}

\question{Who presents the news?} Cable TV news programs feature hosts, anchors
and on-air staff (e.g., contributors, meteorologists) to present the news. We
manually marked 325 of the public figures who we identified in our data set as
news \emph{presenters} (107 on CNN, 130 on FOX, and 88 on MSNBC).  Overall, we
find that a news presenter is on screen 28.1\% of the time -- 27.4\% on CNN,
33.5\% on FOX, and 23.0\% on MSNBC. On CNN, the percentage of time that a news
presenter is on screen increases by 13\% between 2015 and 2018, while it remains
mostly flat over the decade on FOX and MSNBC (\autoref{fig:host_gender_time}a).

The news presenters with the most screen time are Anderson Cooper (1,782~hours) on CNN, Bill
O'Reilly (1,094~h) on FOX, and Rachel Maddow (1,202~h) on MSNBC.  Moreover,
while the top presenter on each channel varies a bit over the course of the
decade (\autoref{fig:top_hosts}), Cooper and O'Reilly hold the top spot for
relatively long stretches on CNN and FOX, respectively. Also, while Maddow appears the
most on MSNBC overall, Chris Matthews holds the top spot for the early part of
the decade (2010 to 2014). Since 2014, the top presenter on MSNBC has
fluctuated on a monthly basis (\autoref{fig:top_hosts}c). The 13\% rise in
screen time of news presenters on CNN that we saw earlier
(in \autoref{fig:host_gender_time}a) can largely be attributed to three hosts
(Anderson Cooper, Chris Cuomo, and Don Lemon),
who see 2.5$\times$, 4.5$\times$, and 5.5$\times$
increases in screen time from 2015 onwards (\autoref{fig:top_hosts}a) and
account for over a third of news presenter screen time on CNN in 2019.

\question{How does screen time of male- and female-presenting news presenters
compare?} The list of top news presenters by screen time is dominated by
male-presenting individuals. Of the top five news presenters on each channel
(accounting for 31\%, 22\%, and 34\% of news presenter screen time on CNN, FOX,
and MSNBC, respectively), only one on CNN and FOX; and two on MSNBC are female
(\autoref{fig:top_hosts}). Across all three channels, there is a shift towards
gender parity in the screen time of news presenters early in the decade followed
by a divergence (\autoref{fig:host_gender_time}b-d).

CNN exhibits gender parity for news presenters in January-June 2012 and
May-August 2015 (\autoref{fig:host_gender_time}b). However, from September 2015
onward, CNN diverges as the 10\% increase in the screen time of male-presenting
news presenters (from 14\% to 24\%) outpaces the 3\% increase for
female-presenting news presenters (13\% to 16\%). The increase in
male-presenting news presenter screen time on CNN mirrors the increase in
overall news presenter screen time on CNN due to an increase in the screen time
for Anderson Cooper, Don Lemon, and Chris Cuomo (\autoref{fig:top_hosts}a).

Similarly, the gender disparity of news presenters on FOX decreases from 2010 to
2016 but widens in 2017 due to an increase in the screen time of male-presenting
news presenters (\autoref{fig:host_gender_time}c). This occurs around the time
of (former top hosts) Megyn Kelly's and Bill O'Reilly's departure from FOX (6\%
and 5\% of presenter screen time, respectively, on FOX in 2016). Their time is
replaced by a rise in Tucker Carlson's and Sean Hannity's screen time (3\% and
5\% of news presenter screen time, respectively, on FOX in 2016 and up to 11\%
and 7\%, respectively, in 2017 and 2018). The increase in female-presenting news
presenter screen time in October 2017 occurs when Laura Ingraham's {\it Ingraham
Angle} and Shannon Bream's {\it FOX News @ Night} debut.

On MSNBC, the disparity as percentage of news presenter screen time increases
from May 2017 to July 2019 (\autoref{fig:host_gender_time}d). This is due to a
similar drop in the screen time of both male- and female-presenting news
presenters. The percentage of time when male-presenting news presenters are on
screen falls from 17\% to 13\%, while the percentage for female-presenting news
presenters falls from 14\% to 7\%. Unlike with CNN and FOX, the change is more
distributed across news presenters; the screen time of the top five presenters
from 2017 to 2019 is comparatively flat (\autoref{fig:top_hosts}c).

\begin{figure}[!tbp]
  \centering
  \includegraphics[width=\columnwidth]{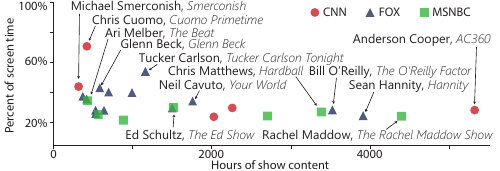}
  \caption{The 25 news presenters who receive the largest fraction of screen time on their own shows (``screenhog''s) and the total amount of video content for their shows in the data set. The top two shows by this metric, {\em Cuomo Primetime} and {\em Tucker Carlson Tonight}, are relatively recent shows, starting in June 2018 and November 2016, respectively.}
  \label{fig:host_screenhogs}
\end{figure}

\question{Which news presenters hog the screen time on their shows?} We compute
the percentage of time a news presenter is on screen on their own show
(``screenhog score'') and plot the top 25 ``screenhog''s
(\autoref{fig:host_screenhogs}). Chris Cuomo (CNN) has the highest fraction of
screen time on his own show (visible 70.6\% of the time on {\it Cuomo
Primetime}). Tucker Carlson (FOX) is second at 55.3\% on {\it Tucker Carlson
Tonight}. These results can be attributed to the format of these two shows;
Cuomo and Carlson both do interviews and often show their own reactions to
guests' comments. Carlson also regularly monologues while on screen. Compared to
both CNN and MSNBC, FOX has the most screenhogs (13 of the top 25), many of whom
are well-known hosts of FOX's opinion shows.  Bill O'Reilly, Anderson Cooper,
and Rachel Maddow (the top presenters by channel) also break the top 25, with
screenhog scores of 28.5\%, 28.3\%, and 24.2\%, respectively.

\begin{figure*}[!tbp]
    \centering
    \includegraphics[width=\textwidth]{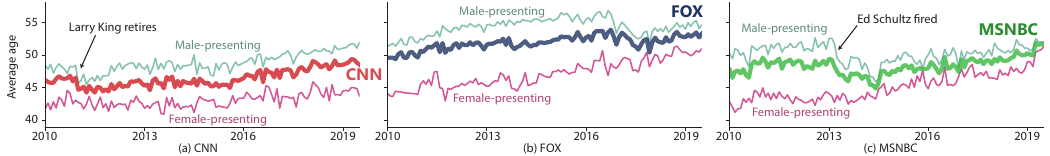}
    \caption{The average age of news presenters, weighted by screen time, has increased on all three channels (bold lines). FOX has the highest average age for both male- and female-presenting news presenters.}
    \label{fig:host_age_increase}
\end{figure*}

\begin{figure}[!tbp]
    \centering
    \includegraphics[width=\columnwidth]{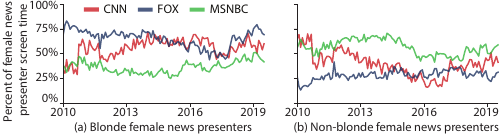}
    \caption{Blonde female news presenters consistently receive more screen time on FOX than non-blonde female news presenters. CNN catches up to FOX from 2014 onward, while the screen time of blonde female news presenters has risen on MSNBC since 2015. On MSNBC, blonde female news presenters do not receive more screen time than non-blonde female news presenters.
    Because blonde and non-blonde female news presenters can be on screen at the same time, the lines in \textbf{(a)} and \textbf{(b)} can add to over 100\%.}
    \label{fig:female_host_haircolor}
\end{figure}

\question{What is the average age of news presenters?} We obtain birthdates for
our list of news presenters from public web sources and we compute the average
age of news presenters on each channel when they are on screen
(\supplemental{1.8}). From 2010 to 2019, the average age of news presenters
rises from 48.2 to 51.0 years (\autoref{fig:host_age_increase}). This trend is
visible for all three channels, though there are localized reversals that are
often marked by the retirements of older, prominent hosts; for example, the
average news presenter's age on CNN falls slightly after Larry King's retirement
in 2010 at age 76. Across all three channels, female-presenting news presenters
are younger on average than their male-presenting counterparts by 6.3 years.
However, the gap has narrowed in recent years.

\question{Are female-presenting news presenters disproportionately blonde?} We
manually annotated the hair color (blonde, brown, black, other) of 145
female-presenting news presenters and computed the screen time of these groups
(\supplemental{1.9}). We find that blondes account for 64.7\% of
female-presenting news presenter screen time on FOX (compared to 28.8\% for
non-blondes). This gives credence to the stereotype that female-presenting news
presenters on FOX fit a particular aesthetic that includes blonde hair
(advanced, for example, by The Guardian~\cite{foxisblonde}). However, counter to
this stereotype, FOX is not alone; the proportion of blondes on CNN (56.6\%
overall and 58.2\% since 2015, compared to 38.6\% overall for non-blondes) has
risen, and, currently, the chance of seeing a blonde female news presenter is
approximately equal on the two networks (\autoref{fig:female_host_haircolor}).
The screen time of blonde female news presenters is lower on MSNBC (36.6\%),
where non-blonde female news presenters account for 55.7\%. On MSNBC, brown is
the dominant hair color at 40.8\%, but 21.4\% is due to a single brown-haired
host (Rachel Maddow). On all three channels, the percentage of blonde female
news presenters far exceeds the natural rate of blondness in the U.S.
($\approx11$\% according to the Bureau of Labor Statistics~\cite{nlsy79}).

\begin{figure*}[!tbp]
  \centering
  \includegraphics[width=\textwidth]{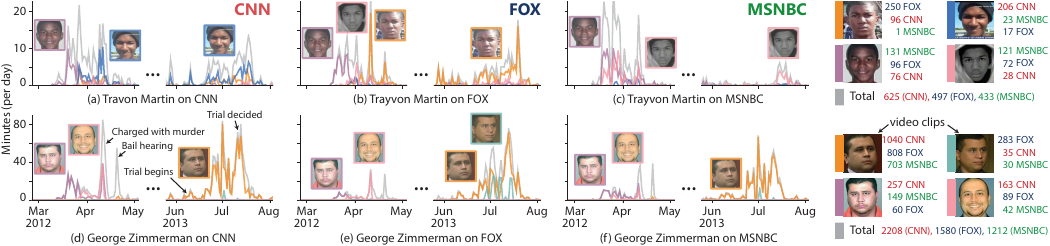}

  \caption{
    In early coverage of the shooting of Trayvon Martin (by George Zimmerman),
    all three channels used the same photos of Martin and Zimmerman. However, as
    the story progressed, depictions of Trayvon (top) differed significantly
    across channels. Depictions of Zimmerman (bottom) also evolved over time,
    but largely reflect efforts by channels to use the most up-to-date photo of
    Zimmerman during criminal proceedings.}

  \label{fig:trayvon_martin_time}
\end{figure*}

\section{How are individuals portrayed?}
\label{sec:portrayal}

Editorial decisions about the images and graphics to include with stories
can subtly influence the way viewers understand a story. We
examine such editorial choices in the context of the Trayvon Martin shooting.

\question{Which photos of Trayvon Martin and George Zimmerman appeared most
often on each channel?} On February 26, 2012, Trayvon Martin, a 17 year-old
high-school student, was fatally shot by neighborhood watchman George
Zimmerman~\cite{trayvonshootingfacts}. Media depictions of both Martin and
Zimmerman were scrutinized heavily as the story captured national
interest~\cite{trayvonphotos,foxtrayvon}. We identified unique photographs of
Martin and Zimmerman in our data set using a K-NN classifier on
FaceNet~\cite{facenet} descriptors and tabulated the screen time of these photos
(see \supplemental{1.10}).

\autoref{fig:trayvon_martin_time} shows the four photos of Martin (top row) and
Zimmerman (bottom row) that received the most screen time in the aftermath of
the shooting and during Zimmerman's 2013 trial. In the initial week of coverage,
all three channels used same image of Martin (purple). This image generated
significant discussion about the ``baby-faced'' depiction of Martin, although it
was dated to a few months before the shooting. In the ensuing weeks (and later
during Zimmerman's trial), differences in how the three channels depict Marin
emerge. CNN most commonly used a photograph of Martin smiling in a blue hat
(blue box). In contrast, the most commonly shown photo on FOX depicts an
unsmiling Martin (orange). MSNBC most frequently used the black-and-white image
of Martin in a hoodie (pink) that was the symbol for protests in support of
Martin and his family. The three different images reflect significant
differences in editorial decisions made by the three channels.

Depictions of Zimmerman also evolved with coverage of the shooting and reflect
both efforts by channels to use the most up-to-date photos for the story at hand
and also the presence of editorial choices. All three channels initially aired
the same image of Zimmerman (purple). The photo, depicting Zimmerman in an
orange polo shirt, was both out of date and taken from a prior police incident
unrelated to the Martin shooting. A more recent photograph of Zimmerman (pink)
was made available to news outlets in late March 2012. While CNN and FOX
transitioned to using this new photo, which depicts a smiling Zimmerman, a
majority of the time, MSNBC continued to give more screen time to the original
photo. After mid-April 2012, depictions of Zimmerman on all three channels
primarily show him in courtroom appearances as the legal process unfolded.

\section{What is discussed in the news?}
\label{sec:text}

The amount of coverage that topics receive in the news can influence viewer
perceptions of world events and newsworthy stories. As a measure of the
frequency of which key topics are discussed, we count the number of times
selected words appear in video captions.

\begin{figure}[!tbp]
    \centering
    \includegraphics[width=\columnwidth]{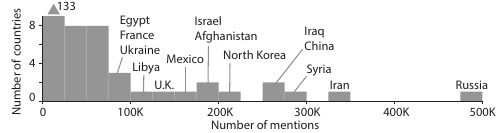}
    \caption{Some countries receive more attention in U.S. cable TV news than others. Russia is the largest outlier followed by Iran.}
    \label{fig:foreign_country_dist}
\end{figure}

\begin{figure*}[!tbp]
    \centering
    \includegraphics[width=\textwidth]{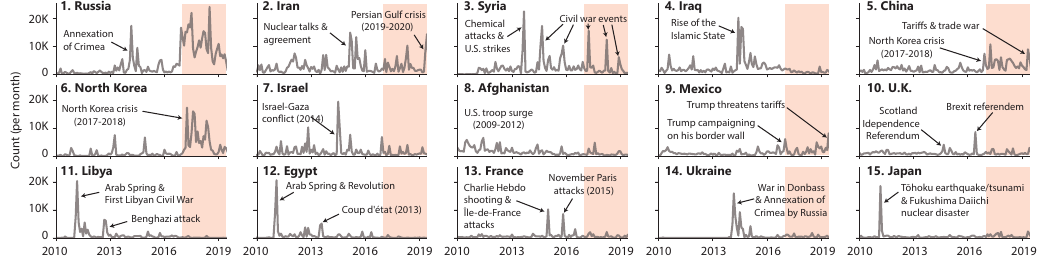}
    \caption{Major peaks in mentions of foreign countries occur around disasters and crises.
    Since the start of Trump's presidency, there has been an increase in coverage of
    Russia, China, and North Korea due to
    increased tensions and a marked shift in U.S. foreign policy (shaded).}
    \label{fig:foreign_country_time}
\end{figure*}

\question{How often are foreign countries mentioned?} Foreign country names,
defined in \supplemental{1.11}, appear in the captions a total of 4.5M times.
Most countries receive little coverage (\autoref{fig:foreign_country_dist}), and
the eight countries with the highest number of mentions (\ngram{Russia},
\ngram{Iran}, \ngram{Syria}, \ngram{Iraq}, \ngram{China}, \ngram{North Korea},
\ngram{Israel}, and \ngram{Afghanistan}) account for 51\% of all country
mentions. \ngram{Russia} alone accounts for 11.2\%. (If treated as a country,
\ngram{ISIS} would rank 2nd after \ngram{Russia} at 8.4\%.) Of these eight, five
have been in a state of armed conflict in the last decade, while the other three
have had major diplomatic rifts with the U.S.. These data suggest that military
conflict and tense U.S. relations beget coverage. No countries from Africa,
South America, and Southeast Asia appear in the top eight; the top countries
from these regions are \ngram{Libya}/\ngram{Egypt} (11th/12th),
\ngram{Venezuela} (32nd), and \ngram{Vietnam} (25th). Bordering the U.S.,
\ngram{Mexico} is 9th, frequently appearing due to disputes over immigration and
trade, while \ngram{Canada} is 21st.

Mentions of individual countries often peak due to important events.
\autoref{fig:foreign_country_time} annotates such events for the 15 most often
mentioned countries. For example, the Libyan Civil War in 2011, the escalation
of the Syrian Civil War in 2012-2013, and the rise of ISIS (Syria, Iraq) in 2014
correspond to peaks. The countries below the top 10 are otherwise rarely in the
news, but the 2011 tsunami and Fukushima Daiichi nuclear disaster; the 2014
annexation of Crimea by Russia; and the Charlie Hebdo shooting and November
Paris attacks (both in 2015), elevated Japan, Ukraine, and France to brief
prominence. Since the election of Donald Trump in 2016, however, there has a
been a marked shift in the top countries, corresponding to topics such as
Russian election interference, North Korean disarmament talks, the Iran nuclear
deal, and the trade war with China.

\begin{figure*}[!tbp]
  \centering
  \includegraphics[width=\textwidth]{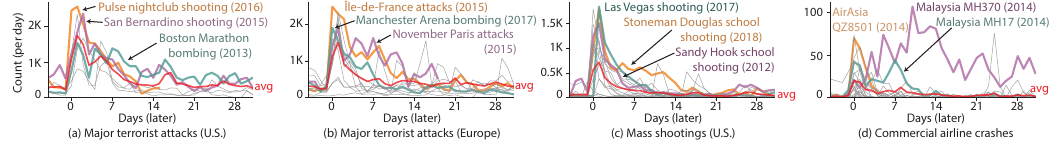}
  \caption{Following a major terrorist attack, mass shooting, or plane crash, usage of related terms increases and remains elevated for 2-3 weeks before returning to pre-event levels. A few plane crashes continued to be covered after this period as new details about the crash (or disappearance in the case of MH370) emerge. In the figure above, lines for individual events are terminated early if another unrelated event of the same category occurs; for example, the San Bernardino shooting (a terrorist attack) in December 2015 occurred three weeks after the November 2015, Paris attacks.}
  \label{fig:how_long_are_events_covered}
\end{figure*}

\question{For how long do channels cover acts of terrorism, mass shootings, and
plane crashes?} We enumerated 18 major terrorist attacks (7 in the U.S. and 11
in Europe), 18 mass shootings, and 25 commercial airline crashes in the last
decade, and we counted related N-grams such as \ngram{terror(ism,ist)},
\ngram{shoot(ing,er)}, and \ngram{plane crash} in the weeks following these
events (\supplemental{1.12} gives the full lists of terms). Counts for
\ngram{terrorism} and \ngram{shootings} return to the pre-event average after
about two weeks (\autoref{fig:how_long_are_events_covered}a-c). Likewise,
coverage of plane crashes also declines to pre-crash levels within two weeks
(\autoref{fig:how_long_are_events_covered}d), though there are some notable
outliers. Malaysia Airlines Flight 370, which disappeared over the Indian Ocean
in 2014, remained in the news for nine weeks, and Malaysia Airlines Flight 17,
shot down over Ukraine, also received coverage for four weeks as more details
emerged.

\begin{figure}[!tbp]
  \centering
  \includegraphics[width=\columnwidth]{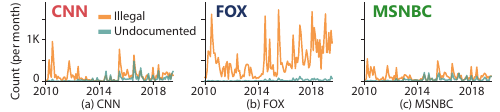}
  \caption{Counts of ``illegal immigrant'' and ``undocumented immigrant'' terminology in video captions, by month. \ngram{Illegal} is more common than \ngram{undocumented} on all three channels, but FOX uses it the most.
  \ngram{Undocumented} only comes into significant use from 2012 onward.}
  \label{fig:immigration}
\end{figure}

\question{Is it illegal or undocumented immigration?} ``Illegal
immigrant'' and ``undocumented immigrant'' are competing terms that describe
individuals who are in the U.S. illegally, with the latter term seen as more
politically correct~\cite{illegalvsundoc}. \autoref{fig:immigration} shows the
counts of when variants of these terms are said (\supplemental{1.13} gives the
full list of variants). \ngram{Illegal} is used on FOX the most (59K times); FOX
also has more mentions of immigration overall. From 2012 onward,
\ngram{undocumented} has increased in use on CNN and MSNBC, though
\ngram{illegal} still appears equally or more often on these channels than
\ngram{undocumented}.

\question{How often are honorifics used to refer to President Trump and Obama?}
Honorifics convey respect for a person or office. We compared the number of
times that \ngram{President (Donald) Trump} is used compared to other mentions
of Trump's person (e.g., \ngram{Donald Trump}, just \ngram{Trump}). When
computing the number of mentions of just \ngram{Trump}, we exclude references to
nouns such as \ngram{the Trump administration} and \ngram{Melania Trump} that
also contain the word Trump, but are not referring to Donald Trump
(\supplemental{1.14} gives the full list of exclusions).

\begin{figure}[!tbp]
  \centering
  \includegraphics[width=\columnwidth]{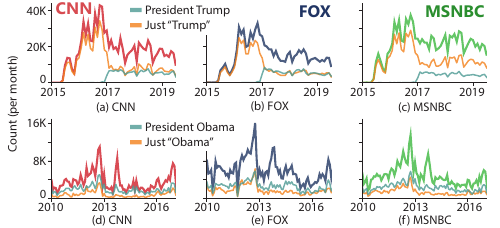}

  \caption{Counts of \ngram{Trump} and \ngram{Obama} peaked in election years
  (2016 and 2012). After his inauguration, \ngram{Trump} is referred to more
  often without \ngram{President} than with (MSNBC has the largest gap). In
  contrast, \ngram{Obama} is referred to with \ngram{President} more often than
  not. The channel color-coded lines represent the total counts of \ngram{Trump}
  and \ngram{Obama}, without exclusions such as \ngram{the Trump administration}.
  Note: most of these counts are captured by the N-grams that we identified
  as references to Trump and Obama's persons.}
  \label{fig:trump_obama_text}
\end{figure}

Our data suggests that although coverage of the incumbent president has
increased since the start of Trump's presidency in 2017, the level of formality
when referring to the president has fallen. Trump, in general, is mentioned
approximately 3$\times$ more than Obama on a monthly basis during the periods of
their respective presidencies in our data set. The term \ngram{President Trump}
only emerges on all three channels following his inauguration to the office in
January 2017 (\autoref{fig:trump_obama_text}a-c). \ngram{President} is used
nearly half of the time on CNN and FOX after his inauguration. By contrast,
MSNBC continues to most commonly refer to him without \ngram{President}. We plot
similar charts of \ngram{President Obama} over the course of his presidency from
2010 to January 2017 (\autoref{fig:trump_obama_text}d-e) and find that, on all
three channels, \ngram{President} is used more often than not.

\section{Who is on screen when a word is said?}
\label{sec:faces_and_text}

People are often associated with specific topics discussed in cable TV news.  We
analyze the visual association of faces to specific topics by computing how
often faces are on screen \emph{at the same time} that specific words are
mentioned. We obtain millisecond-scale time alignments of caption words with the
video's audio track using the Gentle word aligner~\cite{gentlealigner}
(\supplemental{1.1}).

\question{Which words are most likely to be said when women are on screen?} By
treating both face detections and words as time intervals, we compute the
conditional probability of observing at least one female-presenting (or one
male-presenting) face on screen given each word in the caption text
(\supplemental{1.15}). Because of the gender imbalance in screen time, the
conditional probability of a female-presenting face being on screen when
\textit{any} arbitrary word is said is 29.6\%, compared to 61.4\% for a
male-presenting face. We are interested in words where the difference between
female and male conditional probabilities deviates from the baseline 31.9\%
difference.

\begin{figure}[!tbp]
  \centering
  \includegraphics[width=\columnwidth]{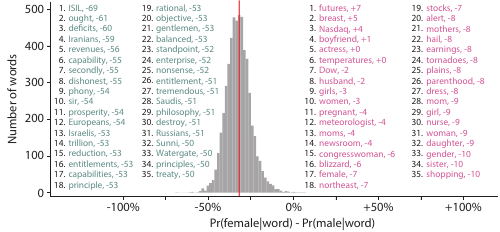}
  \caption{The distribution of words by
  difference in conditional probability of a female- versus a male-presenting
  face being on screen (\supplemental{1.15}).
  The 35 words that are most associated with male- and female-presenting
  screen time are annotated. Note the stark differences in
  topic representation in the top male and female associated words:
  foreign policy, conflict, and fiscal terms (male); and
  female health, family, weather, and business news terms (female).
  }
  \label{fig:female_male_words}
\end{figure}

\autoref{fig:female_male_words} shows the top 35 words most associated with
male- and female-presenting faces on screen. For female-presenting faces, the
top words are about womens' health (e.g., \ngram{breast}, \ngram{pregnant});
family (e.g., \ngram{boyfriend}, \ngram{husband}, \ngram{mom(s)},
\ngram{mothers}, \ngram{parenthood}, etc.); and female job titles (e.g.,
\ngram{actress}, \ngram{congresswoman}). Weather-related terms (e.g.,
\ngram{temperatures}, \ngram{meteorologist}, \ngram{blizzard},
\ngram{tornadoes}) and business news terms (e.g., \ngram{futures},
\ngram{Nasdaq}, \ngram{stocks}, \ngram{earnings}) are also at or near gender
parity, and we attribute this to a number of prominent female weatherpersons
(Indra Petersons/CNN, Janice Dean/FOX, Maria Molina/FOX) and female business
correspondents (Christine Romans/CNN, Alison Kosik/CNN, JJ Ramberg/MSNBC,
Stephanie Ruhle/MSNBC, Maria Bartiromo/FOX) across much of the last decade. The
top words associated with male-presenting faces on screen are about foreign
affairs, terrorism, and conflict (e.g., \ngram{ISIL}, \ngram{Israelis},
\ngram{Iranians}, \ngram{Saudis}, \ngram{Russians}, \ngram{destroy},
\ngram{treaty}); and with fiscal policy (e.g., \ngram{deficits},
\ngram{trillion}, \ngram{entitlement(s)}). The stark difference in the words
associated with female-presenting screen time suggests that, over the last
decade, the subjects discussed on-air by presenters and guests varied strongly
depending on their gender.

\question{Who uses unique words?} We define vocabulary to be ``unique'' to a
person if the probability of that individual being on screen conditioned on the
word being said (at the same time) is high. \autoref{tab:top_individual_words} lists all words for which
an individual has a greater than a 50\% chance of being on screen when the word
is said. (We limit analysis to words mentioned at least 100 times.) Political
opinion show hosts (on FOX and MSNBC) take the most creative liberty in their
words, accounting for all but three names in the list.

\begin{table}[!tbp]
  \centering
  {\small
    \begin{tabularx}{\columnwidth}{ll}
      \textbf{Person} & \textbf{Unique words ($Pr[person|word]$)} \\
      \hline
      Bill O'Reilly (FOX) & opine (60.6), reportage (59.0), \\
        & spout (58.6), urchins (57.9), \\
        & pinhead[ed,s] (49.0, 51.5, 50.2) \\
      Ed Schultz (MSNBC) & classers (71.2), beckster (61.6), \\
        & drugster (59.9), righties (55.2), \\
        & trenders (60.8), psychotalk (54.2) \\
      Tucker Carlson (FOX) & pomposity (76.2), smugness (71.5), \\
        & groupthink (70.5) \\
      Sean Hannity (FOX) & abusively (76.1), Obamamania (53.3) \\
      Glenn Beck (FOX) & Bernays (82.3), Weimar (62.2) \\
      Rachel Maddow (MSNBC) & [bull]pucky (47.9, 50.7), debunktion (51.4) \\
      Chris Matthews (MSNBC) & rushbo (50.5) \\
      \hline
      Kevin McCarthy (politician) & untrustable (75.9) \\
      Chris Coons (politician) & Delawareans (63.8) \\
      Hillary Clinton (politician) & generalistic (56.5) \\
    \end{tabularx}
  }
  \caption{Unique words are often euphemisms or insults (\ngram{urchins}
  $\equiv$ children, \ngram{beckster} $\equiv$ Glenn Beck,
  \ngram{drugster}/\ngram{rushbo} $\equiv$ Rush Limbaugh, \ngram{righties}
  $\equiv$ conservatives, etc.). Others are the names of show segments or
  slogans. For example, \ngram{Psychotalk} is a segment of the {\it Ed Show};
  Sean Hannity refers to the liberal media as \ngram{Obamamania media}; and
  Tucker Carlson brands his own show as the ``sworn enemy'' of \ngram{lying},
  \ngram{pomposity}, \ngram{smugness}, and \ngram{groupthink}.
  Some rare words become unique due to being replayed often on the news; for
  example, Kevin McCarthy (U.S. representative) calls Hillary Clinton
  \ngram{untrustable} and Hillary Clinton uses \ngram{generalistic} in the same
  sentence as her infamous statement characterizing Trump's supporters as
  a ``basket of deplorables''.}
  \label{tab:top_individual_words}
\end{table}

\begin{figure}[!tbp]
  \centering
  \includegraphics[width=\columnwidth]{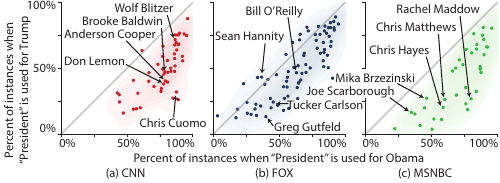}
  \caption{Percentage of mentions that use the \ngram{president} honorific for Trump (post-inauguration to January 20, 2017) and Obama (before January 20, 2017) by each news presenter (dots).
  A majority of presenters on all three channels use \ngram{president} a higher fraction of time when mentioning Obama than they do with Trump. The presenters with the highest screen time on each channel are annotated.}
  \label{fig:hosts_trump_obama}
\end{figure}

\begin{figure*}[!tbp]
  \centering
  \includegraphics[width=\textwidth]{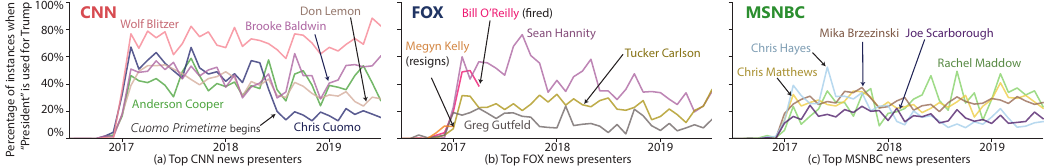}
  \caption{Percentage of time when the \ngram{president} honorific is said for \ngram{Trump}
  while a news presenter is on screen increases after Trump's inauguration
  (top 5 presenters for each channel are shown).
  Chris Cuomo (CNN) drops from over 40\% to under 20\% in June
  2018 with his transition from hosting {\it New Day} to {\it Cuomo Primetime}.
  Sean Hannity's (FOX) decline is more gradual over the course of Trump's
  presidency. From 2017 onward, Wolf Blitzer (CNN) is consistently above the other
  top hosts on any of the three channels (averaging 72\%).}
  \label{fig:top_hosts_trump_time}
\end{figure*}

\begin{figure}[!tbp]
  \centering
  \includegraphics[width=\columnwidth]{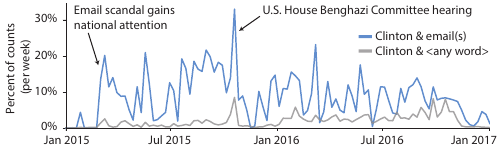}
  \caption{Hillary Clinton is on screen up to 33\% of the time when
  \ngram{email(s)} is mentioned (11\% on average from 2015 to 2016). This is
  significantly higher than the percentage of time that Clinton is on screen
  when any arbitrary word is said (1.9\% on average in the same time period).}
  \label{fig:clinton_email}
\end{figure}

\question{Which presenters are on screen when the President honorific is said?}
A news presenter's use of the \ngram{President} honorific preceding Trump or
Obama might set a show's tone for how these leaders are portrayed.  When a
presenter is on screen, we find that the honorific term \ngram{President} is
used a greater percentage of time for Obama than for Trump, during the periods
of their presidencies. On all three channels, most presenters lie below the
parity line in \autoref{fig:hosts_trump_obama}. However, the average FOX
presenter is closer to parity than the average presenter on CNN or MSNBC in uses
of \ngram{President} in reference to Trump and Obama (a few FOX presenters lie
above the line). Furthermore, \autoref{fig:top_hosts_trump_time} shows how the
top hosts (by screen time) on each channel are associated with uses of
\ngram{President} to refer to Trump over time.

\question{How much was Hillary Clinton's face associated with the word
\ngram{email}?} Hillary Clinton's emails were a frequent news topic in 2015 and
during the 2016 presidential election due to investigations of the 2012 Benghazi
attack and her controversial use of a private email server while serving as U.S.
Secretary of State. During this period, Clinton's face was often on screen when
these controversies were discussed, visually linking her to the controversy. We
compute that during the period spanning 2015 to 2016, Clinton's face is on
screen during 11\% of mentions of the word \ngram{email(s)}
(\autoref{fig:clinton_email}), a significantly higher percentage than the 1.9\%
of the time that she is on screen overall. This degree of association is similar
across all three channels (\supplemental{2.3.1}).

\section{Interactive Visualization Tool}
\label{sec:tool}

\begin{figure*}[t]
    \centering
    \includegraphics[width=\textwidth]{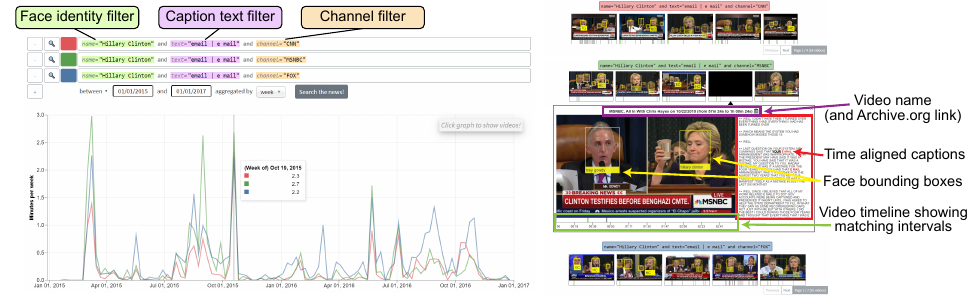}

    \caption{Our interactive visualization tool supports time-series analysis of
    the cable TV news data set. \textbf{(Left)} Users define queries using a combination of
    face, caption text, and video metadata filters. The tool generates
    time-series plots of the total amount of video (aggregate screen time)
    matching these queries. \textbf{(Right)} To provide more context for the
    segments of video included in the chart, users can click on the chart to
    bring up the videos matching the query. We have found that
    providing direct access to the videos is often essential for debugging
    queries and better understanding the relevant video clips.}
    \label{fig:screenshot}

    \vspace{\baselineskip}
    \includegraphics[width=\textwidth]{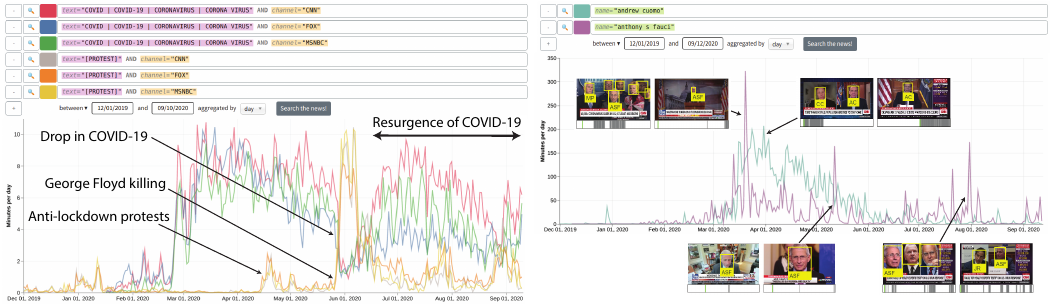}

    \caption{
    Our tool updates daily with new data and can be used to study contemporary
    issues.
    \textbf{(Left)} The amounts of time since December 1, 2019
    when the words \ngram{COVID-19} (and its synonyms) and variants of the
    root word \ngram{PROTEST} are said on each channel,
    treating each utterance as a 1s interval.
    The virus first comes to attention on national cable TV on January
    17, 2020 and peaks on March 12. There is a sharp dip in \ngram{COVID-19}
    (concurrent with a spike in \ngram{PROTEST}) on May 29, when nationwide,
    Black Lives Matter protests following George Floyd's killing took over
    the headlines. From mid-June onward, \ngram{COVID-19}
    coverage rose again; however, the time on FOX is only half that of CNN.
    \textbf{(Right)} New York governor Andrew Cuomo (blue) rose to prominence
    in March-May, but has since disappeared from cable TV,
    while Dr. Anthony S. Fauci (purple) has seen a
    resurgence in screen time since June.}

    \label{fig:covid}
\end{figure*}

We have developed an interactive, web-based visualization tool (available at
\url{https://tvnews.stanford.edu}) that enables the general public to perform
analyses of the cable TV news data set (\autoref{fig:screenshot}). Although this
paper has focused on a static slice of data from January 2010 to July 2019, our
public tool ingests new video daily and can be used to investigate coverage of
contemporary topics (\autoref{fig:covid}). Our design, inspired by the Google
N-gram Viewer~\cite{ngrams}, generates time-series line charts of the amount of
cable TV news video (aggregate time) matching user-specified queries. Queries
may consist of one or more filters that select intervals of time when a specific
individual appears on screen (\texttt{name="..."}), an on screen face has a
specific presented gender (\texttt{tag="male"}), a keyword or phrase appears in
the video captions (\texttt{text="..."}), or the videos come from a particular
channel (\texttt{channel="CNN"}), program, or time of day.

To construct more complex analyses, the tool supports queries containing
conjunctions and disjunctions of filters, which serve to intersect or union the
video time intervals matched by individual filters (\texttt{name="Hillary
Clinton" AND text="email" AND channel="FOX"}).  We implemented a custom
in-memory query processing system to execute screen time aggregation queries
over the entire cable TV news data set while maintaining interactive response
times for the user.  In addition to generating time-series plots of video time,
the tool enables users to directly view underlying video clips (and their
associated captions) that match queries by clicking on the chart.

A major challenge when developing this tool was making an easy-to-use, broadly
accessible data analysis interface, while still exposing sufficient
functionality to support a wide range of analyses of who and what appears on
cable TV news.  We call out three design decisions made during tool development.

\question{(1) Limit visualization to time-series plots.}
Time-series analysis is a powerful way to discover and observe patterns over the
decade spanned by the cable TV news data set. While time-series analysis does
not encompass the full breadth of analyses presented in this paper, we chose to
focus the visualization tool's design on the creation of time-series plots to
encourage and simplify this important form of analysis.

\question{(2) Use screen time as a metric.}
We constrain all queries, regardless of whether visual filters or caption text
filters are used, to generate counts of a single metric: the amount of screen
time matching the query. While alternative metrics, such as using word counts to
analyze of caption text (\autoref{sec:text}) or counts of distinct individuals
to understand who appears on a show, may be preferred for certain analyses, we
chose screen time because it is well suited to many analyses focused on
understanding representation in the news.  For example, a count of a face's
screen time directly reflects the chance a viewer will see a face when turning
on cable TV news. Also, word counts can be converted into screen time intervals
by attributing each instance of a word, regardless of its actual temporal
extent, to a fixed interval of time (\texttt{textwindow="..."}). As a result,
our tool can be used to perform comparisons of word counts as well.

Our decision to make all filters select temporal extents simplified the query
system interface.  All filters result in a selection of time intervals, allowing
all filters to be arbitrarily composed in queries that combine information from
face identity labels and captions.  A system where some filters yielded word
counts and others yields time intervals would complicate the user experience as
it introduces the notion of different data types into queries.

\question{(3) Facilitate inspection of source video clips.}
We found it important for the visualization tool to support user inspection of
the source video clips that match a query (\autoref{fig:screenshot}).
Video clip inspection allows a user to observe the context in which a face or
word appears in a video. This context in turn is helpful for understanding why a
clip was included in a query result, which facilitates deeper understanding of
trends being investigated, aids the process of debugging and refining queries,
and helps a user assess the accuracy of the automatically generated video labels
relied on by a query.

\subsection*{Analysis of user-generated charts}

We released the tool on August 19, 2020 and began analyzing user behavior from
August 27, 2020 onward. As of September 10, 2020, we have logged 2.6K unique
users (based on IP addresses, excluding the authors), who have, on average,
created 1.6 new charts containing one or more queries. We provide a FAQ and
example queries, and these account for 12\% of user-generated charts, while a
further 36\% are modifications to our examples. Of user-generated charts, 43\%
plot the screen time of public figures, 3.7\% plot screen time by gender, and
59\% plot caption text searches (6.7\% are multimodal, with both faces and
text). Excluding names featured in our examples (e.g., Joe Biden, Donald Trump,
Hillary Clinton, Kamala Harris, Elizabeth Warren), the most-queried individuals
are Bernie Sanders, Nancy Pelosi, Barack Obama, Mike Pence, along with several
other 2020 Democratic presidential candidates. Underscoring the value of timely
data, users show interest in current events; many common words are related to
political polarization (e.g., \ngram{QAnon}, \ngram{antifa}, \ngram{postal
service}), COVID-19 (e.g., \ngram{mask(s)}), civil unrest (e.g., \ngram{George
Floyd}, \ngram{protest(s)}, \ngram{looting}), the economy (e.g.,
\ngram{economy}, \ngram{(un)employment}), and technology (e.g., \ngram{AI},
\ngram{computer science}). We hope that allowing the public to analyze such
content will improve media understanding.

\section{Limitations and Discussion}
\label{sec:discussion}

Annotating video using machine learning techniques enables analysis at scale,
but it also presents challenges due to the limitations of automated methods.
Most importantly, the labels generated by computational models have errors, and
understanding the prevalence and nature of labeling errors (including forms of
bias~\cite{facedetectionaudit}) is important to building trust in analysis
results. Labeling errors also have the potential to harm individuals that appear
in cable TV news, in particular when related to gender or
race~\cite{excavatingai,imagenetbias,buolamwini:2018:gendershades}. As a step
toward understanding the accuracy of labels, we validated the output of our face
and commercial detection; presented gender estimation; and person identification
models (for a small subset of individuals) against human-provided labels on a
small collection of frames.  The details of this validation process and the
measured accuracies of models are provided in the supplemental material.

Despite errors in our computational labeling methods at the individual level,
aggregate data about gender representation over time on cable TV news is useful
for understanding gender disparities. Many questions about representation in
cable TV news media similarly concern the subject of race, but we are unaware of
any computational model that can accurately estimate an individual's race from
their appearance (models we have seen have much lower accuracy than models for
estimating presented gender). However, it may be possible to automatically
determine the race of individuals for whom we have an identity label by using
external data sources to obtain the individual's self-reported race.  A similar
procedure could also be used to obtain the self-identified gender of an
individual, reducing our reliance on estimating presented gender from
appearance. Such approaches could further improve our understanding of race and
gender in cable TV news.

Our system lacks mechanisms for automatically differentiating different types of
face appearances.  For example, an individual's face may be on screen because
they are included in infographics, directly appearing on the program (as a
contributor or guest), or shown in B-roll footage.  The ability to differentiate
these cases would enable new analyses of how the news covers individuals.
Likewise, while our query system can determine when a specific individual's face
is on screen when a word is spoken, it does not perform automatic speaker
identification.  As a result, the on screen face may not be speaking -- e.g.,
when a news presenter delivers narration over silent B-roll footage.  Extending
our system to perform automatic speaker
identification~\cite{Ephrat:2018:LookToListen} would allow it to directly
support questions about the speaking time of individuals in news programs or
about which individuals spoke about what stories.

We believe that adding the ability to identify duplicate clips in the data set
would prove to be useful in future analyses.  For example, duplicate clips can
signal re-airing of programs or replaying of popular sound bites. We would also
like to connect analyses with additional data sources such as political
candidate polling statistics~\cite{fivethirtyeight} as well as the number and
demographics of viewers~\cite{nielsen}.  Joining in this data would enable
analysis of how cable TV news impacts politics and viewers more generally.
Finally, we are working with several TV news organizations to deploy private
versions of our tool on their internal video archives.

\section{Related Work}
\label{sec:related}

Prior work in HCI and CSCW has investigated the ``information
environments''\,\cite{Kay2015} created by technologies such as search
engines\,\cite{Hussein2020}, social media feeds\,\cite{Chakraborty2017}, and
online news\,\cite{Diakopoulos2019}. By determining what information is easily
accessible to users, these systems affect people's beliefs about the world. For
example, Kay et. al.\,\cite{Kay2015} showed that gender imbalance in image search results can
reinforce gendered stereotypes about occupations. Common methods
used include algorithmic audits\,\cite{Trielli2019}, mixed-method studies of
online disinformation campaigns\,\cite{Starbird2019}, and user studies that
gauge how algorithms and UI design choices affect user
perceptions\,\cite{Spillane2020, Eslami2015}. While topics such as
misinformation spread via social media and online news have become a popular
area of research in this space, television remains the dominant news format in
the U.S.~\cite{fakenewsecosystem}. Therefore, analysis of traditional media such
as cable TV is necessary to characterize the information environments that users
navigate.

\question{Manual analysis of news and media.}  There have been many efforts to
study trends in media presentation, ranging from analysis of video editing
choices\,\cite{hallin1992,barnhurst1997,bucy2007,quickerdarker}, coverage of
political candidates\,\cite{taiwanbias}, prevalence of segment formats (e.g.
interviews\,\cite{interviewfrequency}), and representation by race and
gender\,\cite{bbc5050,gmmp,WMC:2017:womeninmedia,MediaMatters:2016:diversity}.
These efforts rely on manual annotation of media, which limits analysis to small
amounts of video (e.g., a few 100s of
hours/soundbites\,\cite{hallin1992,bucy2007}, five Sunday morning news shows in
2015\,\cite{MediaMatters:2016:diversity}) or even to anecdotal observations of a
single journalist\,\cite{foxtrayvon,trumpfreemedia}. The high cost of manual
annotation makes studies at scale rare. The BBC 50:50 Project\,\cite{bbc5050},
which audits gender representation in news media, depends on self-reporting from
newsrooms across the world.  GMMP\,\cite{gmmp} relies on a global network of
hundreds of volunteers to compile a report on gender representation every five
years.  While automated techniques cannot generate the same variety of labels as
human annotators (GMMP requires a volunteer to fill out a three-page form for
stories they annotate\,\cite{gmmp}), annotation at scale using computational
techniques stands to complement these manual efforts.

\question{Automated analysis of media.}
Our work was heavily inspired by the Google N-gram viewer~\cite{ngrams} and
Google Trends~\cite{GOOGtrends}, which demonstrate that automated computational
analysis of word frequency, when performed at scale (on centuries of digitized
books or the world's internet search queries), can serve as a valuable tool for
studying trends in culture. These projects allow the general public to conduct
analyses by creating simple time series visualizations of word frequencies. We
view our work as bringing these ideas to cable TV news video.

Our system is similar to the GDELT AI Television
Explorer~\cite{gdelt_ai_explorer}, which provides a web-based query interface
for caption text and on screen chryon text in the Internet Archive's cable TV
news data set and recently added support for queries for objects appearing on
screen. Our work analyzes nearly the same corpus of source video, but, unlike
GDELT, we label the video with information about the faces on screen. We believe
that information about who is on screen is particularly important in many
analyses of cable TV news media, such as those in this paper.

There is growing interest in using automated computational analysis
of text, images, and videos to facilitate understanding of trends in media and
the world.  This includes mining print news and social media to predict civil
unrest~\cite{embers,embers4y} and forced population
migration~\cite{forcedmigration}; using facial recognition on TV video streams
to build connectivity graphs between politicians~\cite{japanfaces}; using gender
classification to quantify the lack of female representation in Hollywood
films~\cite{geenadavis}; understanding presentation style and motion in ``TED
talk'' videos~\cite{huaminvideo,huaminemoco}; identifying trends in
fashion~\cite{Matzen:2017:StreetStyle,Ginosar:2017:yearbook} from internet
images; and highlighting visual attributes of
cities~\cite{Doersch:2012:Paris,arietta2014city}. These prior works address
socially meaningful questions in other domains but put forward
techniques that may also be of interest in our cable TV data set.

Finally, time series visualizations of word and document frequencies are
commonly used to show changes in patterns of cultural production~\cite{epoch}.
We draw inspiration from advocates of ``distant reading,'' who make use of these
visual representations to allow for insights that are impossible from manual
inspection of document collections~\cite{moretti}.

\question{Alternative approaches for video analysis queries.} A wide variety of
systems exist for interactive video analysis, and existing work in interaction
design has presented other potential approaches to formulating queries over
video data sets. Video Lens~\cite{videolens} demonstrates interactive filtering
using brushing and linking to filter complex spatio-temporal events in baseball
video. The query-by-example approach~\cite{querybyexample} has been used in
image~\cite{cao2010mindfinder, ebaysearch, pinterestsearch, conceptcanvas}, and
sports domains~\cite{sha2016chalkboarding, sha2017fine}. These example-based
techniques are less applicable for our visualization tool, which focuses on
letting users analyze who and what is in cable TV news; typing a person's name
or the keywords in the caption is often easier for users than specifying these
attributes by example. Other works from H\"{o}ferlin, et
al.~\cite{facetedexploration} and Meghdadi, et al.~\cite{activeshotsummary}
propose interactive methods to cluster and visualize object trajectories to
identify rare events of interest in surveillance video. Analyzing motion-based
events (e.g., hand gestures) in TV news is an area of future work.

\section{Conclusion}

We have conducted a quantitative analysis of nearly a decade of U.S. cable TV
news video. Our results demonstrate that automatically-generated video
annotations, such as annotations for when faces are on screen and when words
appear in captions, can facilitate analyses at scale that provide unique insight
into trends in who and what appears in cable TV news. To make analysis of our
data set accessible to the general public, we have created an interactive screen
time visualization tool that allows users to describe video selection queries
and generate time-series plots of screen time. We hope that by making this tool
publicly available, we will encourage further analysis and research into the
presentation of this important form of news media.

\begin{acks}
  This material is based upon work supported by the National Science Foundation
  (NSF) under IIS-1539069 and III-1908727. This work was also supported by
  financial and computing gifts from the Brown Institute for Media Innovation,
  Intel Corporation, Google Cloud, and Amazon Web Services. We thank the
  Internet Archive for providing their data set for academic use. Any opinions,
  findings, and conclusions or recommendations expressed in this material are
  those of the author(s) and do not necessarily reflect the views of the
  sponsors.
\end{acks}

\FloatBarrier

\bibliographystyle{ACM-Reference-Format}
\bibliography{tvnews.bib}

\end{document}

\title{Supplemental Material: \\ Detailed Methodology and Additional Analyses}
\author{James Hong, Will Crichton, Haotian Zhang, Daniel Y. Fu, Jacob Ritchie}
\author{Jeremy Barenholtz, Ben Hannel, Xinwei Yao, Michaela Murray}
\author{Geraldine Moriba, Maneesh Agrawala, Kayvon Fatahalian}
\affiliation{
  \institution{Stanford University}
}

\renewcommand{\shortauthors}{Hong, et al.}

\maketitle
\pagestyle{plain}

\section{The data set and processing}
\label{sec:dataset}

Our static data set consists of 244,038 hours of video, audio, and captions
recorded by the Internet Archive's TV News Archive~\cite{tvnewsarchive} from
January 1, 2010 to July 23, 2019. It is segmented into 215,771 videos, organized
by the date/time of airing and the name of the show. The data set requires
114~terabytes to store, encoded in standard definition (640$\times$360 to
858$\times$480) with the H.264 standard. We use Scanner~\cite{scanner}, a
distributed video processing framework, to decode and process the video.

For the up-to-date data used by our interactive tool, the data set is growing
with time (270,000+ hours as of September, 2020).

\subsection{Captions and time alignment}
\label{sec:captions}

Closed captions are available from the Internet Archive. The captions are all
upper case for the majority of news programming and contain 2.49 billion text
tokens, of which 1.94 million are unique (average token length is 3.82
characters). Not all tokens are words (they include punctuation, numbers,
misspellings, etc.), however. By a random sample of the set of unique tokens, we
estimate that there are 141K unique English words in the data set ($\pm$ 31K at
95\% confidence).

We use the Gentle word aligner~\cite{gentlealigner} to perform sub-second
alignment of words in a video's captions to the video's audio track, assigning
each token a starting and ending time. (The source captions are only coarsely
aligned to the video.) Alignment is considered successful if alignments are
found for 80\% of the words in the captions. By this metric, we are able to
align captions for 92.4\% of the videos. The primary causes of failure for
caption alignment are truncated captions or instances where the captions do not match the
audio content (e.g., due to being attributed to the wrong video). The average
speaking time for a single word after alignment is 219 ms.

While the captions are generally faithful to the words being spoken, we observe
occasional differences between the captions and the audio track. For example,
captions are missing when multiple individuals are speaking (interrupting or
talking over each other). The spelling in the captions also sometimes does not
reflect the standard English spelling of a word; \ngram{email} appears as
\ngram{e mail} and \ngram{Obamacare} appears as \ngram{Obama Care}. When
analyzing these topics in the paper, we account for these spelling/segmentation
variants.

\subsection{Commercial detection}
\label{sec:commercials}

We observe that commercial segments in the data set are often bracketed by
black frames, have captions in mixed/lower case (as opposed to all uppercase for
news content), or are missing caption text entirely. Commercials also do not
contain \ngram{>>} delimiters (for speaker changes). Using these features,
we developed a heuristic algorithm that scans videos for sequences of black
frames (which typically indicate the start and end of commercials) and for video
segments where caption text is either missing or mixed/lower case. The algorithm
is written using Rekall~\cite{rekall}, an API for complex event detection in
video, and is shown in \autoref{list:commercial_detection}. To validate our
commercial detection algorithm, we hand annotated 225 hours of videos with 61.8
hours of commercials. The overall precision and recall of our detector on this
annotated data set are 93.0\% and 96.8\% respectively.

Note: we are unable to detect commercials in 9,796 hours of video (2,713 CNN,
4,614 FOX, and 2,469 MSNBC) because the captions from those videos are
unavailable due to failed alignment or missing from the Internet
Archive~\cite{tvnewsarchive}.

\begin{figure}[htp!]
\begin{lstlisting}
# Commercial Query
caption_words = rekall.ingest(captions, 1D)
histograms = rekall.ingest(database.table("hists"), 1D)
entire_video = rekall.ingest(database.table("video"), 3D)

# Find segments with >> delimiters
captions_with_arrows = caption_words
  .filter(word: '>>' in word)

# Find segments of black frames (where all of the pixels
# are black)
black_frame_segs = histograms
  .filter(i: i.histogram.avg() < 0.01)
  .coalesce(predicate = time_gap < 0.1s, merge = time_span)
  .filter(i: i["t2"] - i["t1"] > 0.5s)

# All segments between black frame segments in the
# video are candidates to be considered.
candidate_segs = entire_video.minus(black_frame_seqs)

# Candidate segments that contain >> delimiters are
# rejected
non_commercial_segs = candidate_segs
  .filter_against(
    captions_with_arrows,
    predicate = time_overlaps)

# Keep segments that were not rejected
commercial_segs = entire_video
  .minus(non_commercial_segs.union(black_frame_segs))

# Coalesce any overlapping intervals and filter intervals
# that are too short to be commercials
commercials = commercial_segs
  .coalesce(predicate = time_overlaps, merge = time_span)
  .filter(i: i["t2"] - i["t1"] > 10s)

# Find segments that have lowercase captions
lower_case_word_segs = caption_words
  .filter(word: word.is_lowercase())
  .coalesce(predicate = time_gap < 5s, merge = time_span)

# Find segments that have no captions
no_captions_segs = entire_video
  .minus(caption_words)
  .filter(i: 30 < i["t2"] - i["t1"] < 270)

# Compute the final commercial segments, coalesce
# nearby segments, and reject segments that are too long
commercials = commercials
  .union(lower_case_word_segs)
  .union(no_captions_segs)
  .coalesce(predicate = time_gap < 45s, merge = time_span)
  .filter(comm: comm["t2"] - comm["t1"] < 300s)
\end{lstlisting}
\caption{The Rekall\,\cite{rekall} query for detecting commercials in a video.}
\label{list:commercial_detection}
\end{figure}

\subsection{Face detection}
\label{sub:face_detection}

We use MTCNN~\cite{mtcnn} to detect faces in a subset of frames uniformly spaced
by three seconds in a video. (Performing face detection on all frames is cost
prohibitive.) Three seconds is on the order of 2x the average shot length
($\approx6.2$ seconds between camera cuts) that we estimated for news content
using a shot detection heuristic that checks for large differences in color
histograms between frames. At this sample rate, we detect 306M faces in total,
of which 263M lie in non-commercial video frames. For each of the faces
detected, we compute a 128-dimensional FaceNet~\cite{facenet} descriptor from
the pixels contained within the face's bounding box. These descriptors are used
to compute additional annotations such as binary gender presentation
(\ref{sub:gender_classification}) and person identification for our
self-trained models (\ref{sub:person_identification}).

\begin{table}[!tbp]
  \centering
  \begin{tabular}{crrr}
    \textbf{Year} & \textbf{Precision} & \textbf{Recall} & \textbf{Error (frame level)} \\
    \hline
    2010  & 0.973 & 0.813 & 10.8\% \\
    2011  & 0.986 & 0.792 & 11.2\% \\
    2012  & 0.982 & 0.759 & 14.0\% \\
    2013  & 0.992 & 0.721 & 15.2\% \\
    2014  & 0.979 & 0.757 & 12.8\% \\
    2015  & 0.974 & 0.803 & 11.6\% \\
    2016  & 0.981 & 0.673 & 14.8\% \\
    2017  & 0.984 & 0.720 & 15.2\% \\
    2018  & 0.986 & 0.712 & 17.6\% \\
    2019  & 0.985 & 0.715 & 15.6\% \\
    \hline
    All & 0.985 & 0.745 & 13.9\%
  \end{tabular}
  \caption{Face detector precision and recall for all faces (in 250 randomly sampled frames per year).}
  \label{tab:precision_recall_faces}
\end{table}

To estimate the accuracy of face detection, we manually counted the actual
number of faces and the number of errors (false positives/negatives) made by the
MTCNN~\cite{mtcnn} face detector in 250 randomly sampled frames from each year
of the data (\autoref{tab:precision_recall_faces}). Overall precision is high
($\approx$0.98). Recall is lower ($\approx$0.74) because the metric includes
missed detection of any face, including non-important or difficult to detect
faces (e.g., out-of-focus, partially occluded, very small, extreme side-angled
faces); a large fraction of recall errors are in frames with crowds (such as a
political rally) where background faces are small and often partially occluded.
We also report the percentage of frames that contain at least one error (false
positive or false negative), which is on average 14\% across the entire data
set.

\subsection{Gender classification}
\label{sub:gender_classification}

We trained a binary K-NN classifier using the FaceNet~\cite{facenet}
descriptors. For training data, we manually annotated the presented binary
gender of 12,669 faces selected at random from the data set. On 6,000
independently sampled validation examples, the classifier has 97.2\% agreement
with human annotators. \autoref{tab:gender_confusion} shows the confusion matrix
and class imbalance between male-presenting faces and female-presenting faces.

Imbalances in the error behavior of the K-NN model can influence the results of
an analysis (e.g., recall for females, 93.8\%, is lower than males, 98.8\%). At
present, we do not adjust for these imbalances in the paper. One extension to
our analyses would be to incorporate these error predictions into the reported
findings. For example, we detected 72.5M female-presenting and 178.4M
male-presenting faces in all of the news content (28.9\% of faces are female).
Adjusting based on the error rates in \autoref{tab:gender_confusion}, we would
expect 5.0M females to be mislabeled as males and 2.0M males to be mislabeled as
females, resulting in an expected 76.5M female faces and 175.4M male faces. This
shifts the percentage of female faces to 30.4\%. Similar adjustments to other
analyses where data is analyzed across time or slices (e.g., channel, show,
video segments when ``Obama is on screen'') can be devised, subject to
assumptions about the uniformity and independence of model error rates with
respect to slices of the data set or the availability of additional validation
data to compute fine-grained error estimates. We do not, however, know of
closed-form solutions that are consistently applicable to all of our analyses.
These extensions are considered future work, and we focus on salient differences
in the paper, the accuracy statistics reported here in~\ref{sec:dataset}, and on
careful spot-checking of the results (e.g., using the interactive tool) when
model accuracies are concerned.

In randomly sampling 6,000 faces for validation (4,109 labeled male by human
annotators and 1,891 female), we can estimate that female-presenting individuals
comprise 31.5\% ($\pm$1.2\% at 95\% confidence) of the faces in the data set.

\begin{table}[!tbp]
  \centering
  \begin{tabular}{l|cc}
      & \multicolumn{2}{c}{\textbf{K-NN model labels}}\\
    \textbf{Human labels} & Male & Female \\
    \hline
    Male & 4,058 & 51 \\
    Female & 118 & 1,773 \\
  \end{tabular}
  \caption{Presented gender confusion matrix between K-NN model generated labels and human labels. The estimated precision and recall for the male-presenting and female-presenting classes are 97.2\% and 98.8\%; and 97.2\% and 93.8\%, respectively.}
  \label{tab:gender_confusion}
\end{table}

\subsection{Identifying public figures}
\label{sub:person_identification}

To identify individuals, we use the Amazon Rekognition Celebrity Recognition
API\,\cite{amazonrekognition}. This API identifies 46.2\% of the faces in the
data set. To reduce flickering (where a portion of instances of an individual in
a video are missed by Amazon), we propagate these face detections to an
additional 10.7\% of faces using a conservative L2 distance metric threshold
between the FaceNet\,\cite{facenet} descriptors of identified and unidentified
faces within the same video.

As mentioned in the paper 1,260 unique individuals receive at least 10
hours of screen time in our data set, accounting in total for 47\% of faces in
the data set. We validated a stratified sample of these individuals and estimate
that 97.3\% of the individuals in this category correspond to people who are in
data set (not just visually similar ``doppelgangers'' of individuals in the
news). See~\autoref{tab:amazon_validation} for the full statistics and
methodology of the doppelgangers estimation.

\begin{table}[!tbp]
  \centering
  \begin{tabular}{lrr}
    \textbf{Screen time} & \textbf{\# of individuals} & \textbf{Est. \% of doppelgangers} \\
    \hline
    0-10 min & 129,138 & - \\
    10-15 min & 8,559 & 80\% \\
    15-30 min & 10,664 & 76\% \\
    30-60 min & 6,352 & 72\% \\
    1-2 hr & 3,403 & 84\% \\
    2-5 hr & 2,136 & 68\% \\
    5-10 hr & 795 & 52\% \\
    10-20 hr & 445 & 4\% \\
    20-50 hr & 415 & 4\% \\
    50-100 hr & 203 & 0\% \\
    100-200 hr & 90 & 0\% \\
    200 hrs or more & 107 & 0\%
  \end{tabular}
  \caption{Amazon Rekognition Celebrity Recognition~\cite{amazonrekognition} returns facial identity predictions for 162,307 distinct names in our data set. We noticed that the majority of uncommon names (individuals with less than 10 hrs of screen time) predicted by Amazon are ``doppelgangers'' of the people who are actually in the news content (false positives). These doppelgangers include a large number of foreign musicians, sports players, and actors/actresses. To evaluate the effect of these errors, we randomly sampled 25 individuals (by name) from each screen time range and visually validated whether the individual is present only as a doppelganger to other individuals. Our results suggest that a threshold of 10 hours is needed to eliminate most of the doppelgangers. We manually verified that the individuals (e.g., politicans, news presenters, shooting perpetrators/victims) referenced in the paper do not fall under the ``doppelganger'' category.}
  \label{tab:amazon_validation}
\end{table}

For important individuals who are not recognized by the Amazon Rekognition
Celebrity Recognition API~\cite{amazonrekognition} or whose labels are known to
be inaccurate, we train our own person identification models using the
FaceNet\,\cite{facenet} descriptors. In the latter case, we determined a
person's labels to inaccurate if they were consistently being missed or
mis-detected on visual inspection of the videos. To obtain our own labels, we
followed two human-in-the-loop labeling methodologies optimized for people who
are common (e.g., a President or news presenter who appears for hundreds of
hours) and for people who are uncommon (e.g., a shooting victim or less-known
public official). The methodologies are described in \ref{ssub:uncommon_people}
and \ref{ssub:common_people}, respectively. We determined which approach to use
experimentally; if we could not find enough training examples for the common
person approach, we switched to the uncommon person approach. The individuals
for which we use our own labels are listed in~\autoref{tab:own_models}.

\autoref{tab:precision_recall_people} estimates the precision and recall of the
labels for the individuals referenced in our paper analyses (e.g., important
political figures and candidates). Precision is influenced by many
factors, including the presence of individuals of similar appearance being
prominent in the news. Because each individual represents only a small portion
of overall face screen time, unbiased recall is difficult to compute without
finding all instances of an individual. We perform a best effort attempt to
estimate recall by manually counting false negatives in randomly sampled frames
from videos known to contain the individual (25 videos, 100 frames per video).
We note that the number of samples per individual, found in these frames, varies
due to the quantity and nature of an individual's coverage (e.g., appearances in
interviews, and the size and quality of their images).

\subsubsection{Methodology for detecting uncommon individuals}
\label{ssub:uncommon_people}

To detect uncommon individuals (with less than $\approx50$ hours of screen time or
60,000 face detections), we use Google Image Search~\cite{googleimages} to
obtain initial images of the person. Next, we use FaceNet~\cite{facenet} to
compute descriptors on these examples. We compute the L2 distances from
these descriptors to descriptors for all other faces in the data set and display
the faces visually by ascending L2 distance. We select instances of the faces
that visually match the person, add them to the example set and repeat the process of
computing L2 distances and displaying images until it becomes difficult to
find additional examples (the top candidates are all images of other people). To make
the selection process more time-efficient, we implemented range navigation and
selection to label faces between L2 distance ranges at once if all or nearly all
of the faces in the range are the correct person. Even so, the primary
limitation of this approach is that the labeling time scales linearly with the
frequency of the individual in the data set.

\subsubsection{Methodology for detecting common individuals}
\label{ssub:common_people}

To detect common individuals, for whom it is impossible to browse all of their
detections, we trained a simple logistic classifier on the FaceNet~\cite{facenet} features. We
used Google Image Search~\cite{googleimages} to find initial examples and
augment those by sampling faces from the data set that are similar to the
examples in FaceNet descriptor space. For negative examples, we sample faces
randomly and manually inspect the random samples that are most likely (based on
L2 distance) to be positive examples. (This step is necessary because common
individuals such as Donald Trump are likely to appear in the negative samples
due to their high frequency in the data set.) We then use these positive and
negative examples to train a model. To improve the model, we sampled faces for
which the model produces low confidence scores ($\approx0.5$) and labeled these as
new examples, repeating the training and labeling process until finding new
positive examples becomes challenging and model precision is sufficient
(evaluated by visually validating the faces that are labeled positive by the
mode).

\begin{table}[!tbp]
  \centering
  \begin{tabular}{ll}
    \textbf{Politicians} & \textbf{Notes} \\
    \hline
    Donald Trump & Low recall from Amazon \\
    Hillary Clinton & Used for consistency to Trump \\
    Barrack Obama & Used for consistency to Trump \\
    Bernie Sanders & Used for consistency to Trump \\
    Mitt Romney & Used for consistency to Trump \\
    Dick Durbin & Not identified by Amazon \\
    \\
    \textbf{News presenters} \\
    \hline
    Ana Cabrera & Not identified by Amazon \\
    Brian Shactman & Not identified by Amazon \\
    Bryan Illenas & Not identified by Amazon \\
    Dave Briggs & Not identified by Amazon \\
    David Gura & Not identified by Amazon \\
    Dorothy Rabinowitz & Not identified by Amazon \\
    Doug McKelway & Not identified by Amazon \\
    Ed Lavandera & Not identified by Amazon \\
    Griff Jenkins & Not identified by Amazon \\
    Jason Riley & Not identified by Amazon \\
    Jillian Mele & Not identified by Amazon \\
    Jim Pinkerton & Not identified by Amazon \\
    JJ Ramberg & Not identified by Amazon \\
    Lauren Ashburn & Not identified by Amazon \\
    Leland Vittert & Not identified by Amazon \\
    Louis Burgdorf & Not identified by Amazon \\
    Maria Molina & Not identified by Amazon \\
    Natalie Allen & Not identified by Amazon \\
    Nicole Wallace & Not identified by Amazon \\
    Pete Hegseth & Not identified by Amazon \\
    Richard Lui & Not identified by Amazon \\
    Rick Folbaum & Not identified by Amazon \\
    Rick Reichmuth & Not identified by Amazon \\
    Rob Schmitt & Not identified by Amazon \\
    Toure Neblett & Not identified by Amazon \\
    Trace Gallagher & Not identified by Amazon \\
    Yasmin Vossoughian & Not identified by Amazon \\
    \\
    \textbf{Miscellaneous} \\
    \hline
    George Zimmerman & Used for consistency to Martin \\
    Trayvon Martin & Not identified by Amazon \\
  \end{tabular}
  \caption{Individuals for whom we use our own labels. We use our own labels when no labels from Amazon~\cite{amazonrekognition} are available, the Amazon labels are known to have low precision or recall, or to be consistent on major comparisons between individuals labeled with our models and with Amazon.}
  \label{tab:own_models}
\end{table}

\begin{table*}[tpb]
  \centering
  \begin{tabular}{l|rr|rr}
    \multicolumn{1}{l}{\textbf{Name}} &
    \textbf{Samples} & \multicolumn{1}{l}{\textbf{Est. precision}} &
    \textbf{Samples} & \textbf{Est. recall} \\

    \multicolumn{5}{l}{\textbf{U.S. political figures and candidates}} \\
    \hline
    Amy Klobuchar & 100 & 1.00 & 69 & 0.87 \\
    Barack Obama~\textdagger & 100 & 1.00 & 85 & 0.86 \\
    Ben Carson & 100 & 0.99 & 132 & 0.85 \\
    Bernie Sanders~\textdagger & 100 & 0.99 & 42 & 0.83 \\
    Beto O'Rourke & 100 & 1.00 & 50 & 0.58 \\
    Bill Clinton & 100 & 0.89 & 59 & 0.90 \\
    Bill De Blasio & 100 & 1.00 & 55 & 0.89 \\
    Bobby Jindal & 100 & 0.99 & 133 & 1.00 \\
    Carly Fiorina & 100 & 0.92 & 99 & 0.74 \\
    Chris Christie & 100 & 0.98 & 118 & 0.87 \\
    Dick Durbin~\textdagger & 100 & 0.96 & 50 & 0.80 \\
    Donald Trump~\textdagger & 100 & 0.91 & 65 & 0.83 \\
    Elizabeth Warren & 100 & 0.97 & 42 & 0.81 \\
    Gary Johnson & 100 & 0.99 & 124 & 0.84 \\
    George W. Bush & 100 & 0.72 & 71 & 0.80 \\
    Harry Reid & 100 & 0.97 & 137 & 0.83 \\
    Herman Cain & 100 & 1.00 & 100 & 0.90 \\
    Hillary Clinton~\textdagger & 100 & 0.89 & 136 & 0.84 \\
    Jeb Bush & 100 & 0.96 & 79 & 0.92 \\
    Jim Gilmore & 100 & 0.98 & 157 & 0.94 \\
    Jim Webb & 99 & 0.99 & 158 & 0.89 \\
    Joe Biden & 100 & 1.00 & 66 & 0.91 \\
    John Boehner & 100 & 1.00 & 84 & 0.95 \\
    John McCain & 99 & 0.99 & 196 & 0.91 \\
    Jon Huntsman Jr. & 100 & 1.00 & 117 & 0.87 \\
    Kamala Harris & 99 & 0.97 & 55 & 0.93 \\
    Kellyanne Conway & 100 & 1.00 & 151 & 0.72 \\
    Kevin McCarthy & 100 & 1.00 & 70 & 0.97 \\
    Lincoln Chafee & 100 & 0.88 & 103 & 0.87 \\
    Lindsey Graham & 100 & 1.00 & 107 & 0.88 \\
    Marco Rubio & 100 & 1.00 & 93 & 0.85 \\
    Martin O'Malley & 100 & 0.92 & 129 & 0.86 \\
    Michele Bachmann & 100 & 0.91 & 104 & 0.92 \\
    Michelle Obama & 100 & 1.00 & 107 & 0.76 \\
    Mike Huckabee & 100 & 1.00 & 299 & 0.96 \\
    Mitch McConnell & 99 & 1.00 & 81 & 0.83 \\
    Mitt Romney~\textdagger & 100 & 0.98 & 107 & 0.72 \\
    Nancy Pelosi & 100 & 1.00 & 37 & 0.87 \\
    Newt Gingrich & 100 & 0.98 & 226 & 0.94 \\
    Orrin Hatch & 100 & 0.99 & 115 & 0.94 \\
    Paul Ryan & 100 & 0.99 & 104 & 0.84 \\
    Pete Buttigieg & 100 & 0.99 & 25 & 0.96 \\
    Rand Paul & 100 & 1.00 & 140 & 0.94 \\
    Rick Santorum & 100 & 1.00 & 168 & 0.92 \\
    Rick Perry & 100 & 0.99 & 154 & 0.77 \\
    Ron Paul & 100 & 1.00 & 185 & 0.96 \\
    Sarah Palin & 100 & 1.00 & 126 & 0.85 \\
    Steve Scalise & 100 & 0.97 & 109 & 0.94 \\
    Ted Cruz & 100 & 1.00 & 102 & 0.85 \\
    Tim Kaine & 100 & 0.99 & 185 & 0.92 \\
    Tulsi Gabbard & 100 & 0.97 & 88 & 0.78 \\
    \multicolumn{5}{l}{}\\
    \multicolumn{5}{l}{\textbf{Miscellaneous}} \\
    \hline
    George Zimmerman~\textdagger & 100 & 0.98 & 131 & 0.79 \\
    Trayvon Martin~\textdagger & 100 & 0.95 & 48 & 0.63 \\
  \end{tabular}
  \caption{Estimated precision is computed on $\approx$100 randomly
  sampled faces identified as each individual.
  Estimated recall is computed on actual instances of each individual's face
  found in a random sample of 2,500 faces, from 25 videos, known to
  contain each individual.
  (\textdagger~indicates our models.)}
  \label{tab:precision_recall_people}
\end{table*}

\subsection{Enumerating news presenters}
\label{sub:presenters}

TV news networks refer to their hosts and staff members using a number of terms
(e.g., hosts, anchors, correspondents, personalities, journalists); these terms
vary by role and by network. We use the term ``news presenter'' to refer broadly
to anchors, hosts, and on-air staff (contributors, meteorologists, etc.) of a
news network, and we manually enumerated 325 news presenters from the three
networks~\autoref{tab:all_hosts_staff}. Our list of names consists of the staff
listings on the public web pages of CNN, FOX, and MSNBC, accessed in January
2020, and information manually scraped from Wikipedia for the top 150 shows by
screen time (accounting for 96\% of news content). Because content is shared
between different channels at a network, the list for CNN also includes
presenters from HLN, owned by CNN. NBC and CNBC presenters are also included in
the MSNBC list. We were unable to identify faces for 18 presenters and these
individuals are excluded from the 325 presenters listed. These omitted
individuals are either not recognized by Amazon
Rekognition~\cite{amazonrekognition}, not in the video data set (e.g.,
presenters only on HLN or CNBC), or too rare to detect reliably in our data set
(e.g., left before January 1, 2010; joined after July 23, 2019; or very specific
domain experts).

Most presenters are enumerated at the granularity of a channel; Anderson Cooper
(who is a host on CNN) is considered to be a presenter in any CNN video, but
would not be considered a presenter on FOX or MSNBC. We do not differentiate
between presenter roles, and a presenter's role may change over the decade as
they are promoted or move from show to show. We also do not track the exact
length of employment for each presenter on a network; however, the screen time
of presenters on a channel becomes negligible (near zero) after they have left
the network (due to changing employer, retiring, or being fired). Some
presenters in our data set have moved between channels; for example, Ali Velshi
left CNN in 2013 and joined MSNBC in 2016. For individuals who were prominent
political figures before becoming news presenters, we track presenter status at
the show granularity (e.g., Mike Huckabee, Newt Gingrich, and David Axelrod).
\autoref{tab:all_hosts_staff} lists all of the news presenters who we identify.

\begin{table*}[p]
  \vspace{-6em}
  {\small
  \centering
  \begin{tabular}{lllll}
    \textbf{CNN} \\
    \hline
    Ali Velshi (225.9 hours) & Alison Kosik (104.3) & Alisyn Camerota (271.1) & Amanda Davies (3.4) & Amara Walker (9.5) \\
    Ana Cabrera (305.7) & Anderson Cooper (1782.3) & Andrew Levy (0.0) & Anthony Bourdain (110.8) & Arwa Damon (50.1) \\
    Ashleigh Banfield (193.2) & Barbara Starr (156.5) & Becky Anderson (12.2) & Ben Wedeman (61.9) & Bianna Golodryga (16.0) \\
    Bill Hemmer (0.2) & Bill Weir (16.0) & Brian Stelter (188.6) & Brianna Keilar (267.3) & Brooke Baldwin (898.6) \\
    Campbell Brown (28.8) & Candy Crowley (140.7) & Carol Costello (311.4) & Chris Cuomo (678.0) & Christi Paul (84.1) \\
    Christiane Amanpour (72.6) & Christine Romans (315.0) & Clarissa Ward (33.0) & Dana Bash (350.4) & Dave Briggs (91.7) \\
    Deborah Feyerick (80.2) & Don Lemon (1098.8) & Drew Griffin (86.4) & Ed Lavandera (57.0) & Elizabeth Cohen (35.2) \\
    Erica Hill (57.4) & Erin Burnett (539.6) & Errol Barnett (63.5) & Fareed Zakaria (230.3) & Frederik Pleitgen (71.4) \\
    Fredricka Whitfield (477.8) & Gary Tuchman (37.4) & Gloria Borger (255.6) & Hala Gorani (28.6) & Howard Kurtz (39.0) \\
    Jake Tapper (376.3) & Jamie Gangel (17.7) & Jean Casarez (35.5) & Jeff Zeleny (115.2) & Jeffrey Toobin (270.6) \\
    Jessica Yellin (73.1) & Jim Acosta (220.9) & Jim Sciutto (282.3) & Joe Johns (118.3) & John Berman (584.2) \\
    John King (377.0) & John Roberts (46.6) & John Vause (62.2) & John Walsh (20.6) & Kate Bolduan (322.6) \\
    Kathleen Parker (21.7) & Kiran Chetry (54.5) & Kristie Lu Stout (4.2) & Kyra Phillips (105.1) & Kyung Lah (47.9) \\
    Larry King (78.9) & Lisa Ling (25.2) & Lou Dobbs (0.3) & Lynda Kinkade (5.4) & Lynn Smith (0.2) \\
    Martin Savidge (91.7) & Max Foster (34.4) & Michael Smerconish (177.6) & Michelle Kosinski (49.2) & Miguel Marquez (0.2) \\
    Mike Galanos (2.4) & Mike Rogers (50.0) & Mike Rowe (4.8) & Morgan Spurlock (13.3) & Natalie Allen (75.5) \\
    Nic Robertson (135.5) & Nick Paton Walsh (65.3) & Pamela Brown (110.2) & Paula Newton (17.3) & Piers Morgan (404.2) \\
    Poppy Harlow (209.5) & Rachel Nichols (31.6) & Randi Kaye (148.0) & Richard Quest (90.5) & Richard Roth (7.4) \\
    Robin Meade (2.0) & Rosemary Church (81.6) & S. E. Cupp (45.7) & Sanjay Gupta (200.1) & Sara Sidner (21.0) \\
    Soledad O'Brien (91.6) & Stephanie Cutter (14.8) & Susan Hendricks (19.3) & Suzanne Malveaux (130.8) & T. J. Holmes (114.4) \\
    Tom Foreman (44.0) & Van Jones (156.2) & Victor Blackwell (113.8) & W. Kamau Bell (43.9) & Wolf Blitzer (800.1) \\
    Zain Asher (23.4) & Zain Verjee (24.2) &  &  &  \\
    \\

    \textbf{FOX} \\
    \hline
    Abby Huntsman (51.3) & Ainsley Earhardt (211.9) & Alan Colmes (65.3) & Alisyn Camerota (141.3) & Andrea Tantaros (177.5) \\
    Andrew Levy (160.1) & Andrew Napolitano (122.6) & Angela McGlowan (23.7) & Anna Kooiman (78.6) & Ari Fleischer (31.9) \\
    Arthel Neville (108.9) & Bill Hemmer (383.0) & Bill O'Reilly (1093.8) & Bob Beckel (268.1) & Brenda Buttner (34.8) \\
    Bret Baier (536.7) & Brian Kilmeade (638.4) & Brit Hume (171.7) & Bryan Llenas (33.6) & Byron York (77.8) \\
    Cal Thomas (13.9) & Carol Alt (8.9) & Casey Stegall (26.2) & Charles Krauthammer (283.2) & Charles Payne (98.1) \\
    Charlie Gasparino (45.9) & Cheryl Casone (33.1) & Chris Wallace (374.3) & Clayton Morris (217.4) & Dagen McDowell (44.4) \\
    Dana Perino (437.2) & Daniel Henninger (53.1) & Dave Briggs (70.1) & David Asman (50.1) & David Hunt (1.0) \\
    Dorothy Rabinowitz (7.2) & Doug McKelway (68.9) & Ed Henry (313.4) & Ed Rollins (32.1) & Elisabeth Hasselbeck (85.4) \\
    Elizabeth Prann (25.0) & Ellis Henican (6.2) & Eric Bolling (394.9) & Eric Shawn (128.7) & Fred Barnes (10.8) \\
    Geraldo Rivera (232.3) & Gerri Willis (27.8) & Glenn Beck (288.1) & Greg Gutfeld (782.5) & Greta van Susteren (487.5) \\
    Gretchen Carlson (268.1) & Griff Jenkins (31.6) & Guy Benson (52.6) & Harris Faulkner (291.7) & Heather Childers (201.4) \\
    Howard Kurtz (227.0) & James Taranto (4.6) & Jane Hall (0.1) & Janice Dean (41.6) & Jason Riley (25.3) \\
    Jeanine Pirro (514.4) & Jedediah Bila (71.3) & Jehmu Greene (21.3) & Jennifer Griffin (57.9) & Jesse Watters (290.7) \\
    Jillian Mele (118.9) & Jim Pinkerton (24.6) & John Fund (20.1) & John Roberts (65.5) & John Stossel (119.8) \\
    Jon Scott (300.3) & Juan Williams (367.0) & Judith Miller (51.3) & Julie Banderas (98.2) & Karl Rove (252.4) \\
    Katherine Timpf (60.2) & Katie Pavlich (83.4) & Kelly Wright (71.9) & Kevin Corke (40.1) & Kimberley Strassel (56.0) \\
    Kimberly Guilfoyle (258.7) & Kristen Soltis Anderson (10.1) & Laura Ingle (31.0) & Laura Ingraham (498.0) & Lauren Ashburn (9.5) \\
    Lauren Green (8.6) & Leland Vittert (136.6) & Leslie Marshall (73.3) & Manny Alvarez (12.2) & Mara Liasson (25.5) \\
    Maria Bartiromo (81.0) & Maria Molina (67.1) & Mark Fuhrman (29.1) & Mark Levin (55.4) & Martha Maccallum (562.5) \\
    Megyn Kelly (790.9) & Melissa Francis (84.6) & Michael Baden (19.5) & Mike Emanuel (99.7) & Molly Henneberg (28.4) \\
    Molly Line (30.7) & Monica Crowley (89.3) & Neil Cavuto (737.6) & Paul Gigot (98.7) & Pete Hegseth (246.6) \\
    Peter Doocy (87.9) & Phil Keating (35.9) & Rachel Campos-Duffy (11.2) & Raymond Arroyo (21.9) & Rich Lowry (37.0) \\
    Rick Folbaum (42.8) & Rick Reichmuth (80.7) & Rob Schmitt (51.0) & Robert Jeffress (19.4) & Sandra Smith (71.7) \\
    Sean Hannity (1071.8) & Shannon Bream (416.0) & Shepard Smith (360.2) & Steve Doocy (450.4) & Steve Hilton (81.2) \\
    Stuart Varney (126.2) & Tammy Bruce (60.5) & Tom Shillue (145.5) & Tomi Lahren (15.8) & Trace Gallagher (131.7) \\
    Trish Regan (44.5) & Tucker Carlson (865.3) & Uma Pemmaraju (48.3) & Walid Phares (28.2) & William Bennett (16.9) \\
    \\

    \textbf{MSNBC} \\
    \hline
    Abby Huntsman (29.0) & Al Sharpton (286.7) & Alec Baldwin (2.5) & Alex Wagner (174.8) & Alex Witt (261.3) \\
    Ali Velshi (242.4) & Andrea Canning (4.9) & Andrea Mitchell (392.2) & Andrew Ross Sorkin (11.0) & Angie Goff (1.3) \\
    Anne Thompson (10.1) & Ari Melber (395.0) & Ayman Mohyeldin (150.4) & Betty Nguyen (29.5) & Bill Neely (20.6) \\
    Brian Shactman (213.3) & Brian Sullivan (13.3) & Brian Williams (282.5) & Carl Quintanilla (0.8) & Chris Hayes (839.5) \\
    Chris Jansing (254.8) & Chris Matthews (1103.8) & Chuck Todd (550.3) & Contessa Brewer (49.8) & Craig Melvin (173.4) \\
    David Faber (1.2) & David Gura (54.9) & Donny Deutsch (53.3) & Dylan Ratigan (109.7) & Ed Schultz (493.0) \\
    Frances Rivera (44.2) & Greta van Susteren (21.7) & Hallie Jackson (105.0) & Jim Cramer (8.0) & Jj Ramberg (30.8) \\
    Joe Scarborough (940.4) & John Heilemann (147.1) & Jose Diaz-Balart (88.3) & Josh Mankiewicz (13.1) & Joy-Ann Reid (337.1) \\
    Kasie Hunt (112.6) & Kate Snow (51.6) & Katy Tur (187.1) & Kayla Tausche (2.1) & Keith Olbermann (109.7) \\
    Kelly Evans (0.7) & Kelly O'Donnell (57.2) & Kerry Sanders (25.2) & Kristen Welker (212.2) & Krystal Ball (91.0) \\
    Lawrence O'Donnell (688.0) & Lester Holt (13.1) & Louis Burgdorf (29.8) & Lynn Smith (28.0) & Mara Schiavocampo (18.5) \\
    Mark Halperin (158.9) & Martin Bashir (114.7) & Matt Lauer (8.4) & Melissa Harris-Perry (197.9) & Meredith Vieira (1.2) \\
    Miguel Almaguer (9.3) & Mika Brzezinski (696.7) & Mike Viqueira (46.9) & Natalie Morales (4.7) & Nicole Wallace (175.9) \\
    Pete Williams (105.4) & Peter Alexander (97.5) & Rachel Maddow (1201.7) & Rehema Ellis (7.2) & Richard Engel (114.2) \\
    Richard Lui (146.4) & Rick Santelli (1.3) & Ron Mott (16.7) & Ronan Farrow (31.4) & Savannah Guthrie (43.9) \\
    Seema Mody (1.5) & Stephanie Gosk (14.0) & Stephanie Ruhle (111.5) & Steve Kornacki (358.6) & Steve Liesman (4.7) \\
    Sue Herera (1.8) & Tamron Hall (200.5) & Thomas Roberts (198.8) & Tom Brokaw (29.2) & Tom Costello (24.5) \\
    Toure Neblett (65.4) & Willie Geist (319.3) & Yasmin Vossoughian (66.6) &  &  \\
  \end{tabular}
  \caption{Compiled list of news presenters and their screen time in hours. Note that the percentage of
    female-presenters in the news presenter list is 52\%, 42\%, and 44\% on CNN, FOX, and MSNBC, respectively.}
  \label{tab:all_hosts_staff}}
\end{table*}

\subsection{Computing ``screenhog score'' for presenters}

``Screenhog score'' is defined in the paper as the percentage of time that a
news presenter is on screen in the content portion of their own show. We
considered shows with at least 100 hours of news content when listing the top 25
news presenters by their screenhog score.

\subsection{Age for news presenters}

We successfully obtained birthdates for 98\% of news presenters using
DBpedia~\cite{dbpedia} and manual Google and Wikipedia~\cite{wikipedia} search.
For the birthdates queried from DBpedia, we manually verified the results to
eliminate common errors such as the wrong birthdate due to the existence of
another person of the same name. In a small number of cases (1\%), only the
birth year was available; for these individuals, we compute their age from
January 1 of their birth year.

We calculate the age of news presenters, weighted by screen time, by
assigning each face identified as a news presenter with the age (at day
granularity) of the individual on the day that the video aired. The average age,
weighted by screen-time corresponds to the expected age of a news presenter
sampled randomly.

\begin{figure}[!tbp]
  \centering
  \includegraphics[width=0.32\columnwidth]{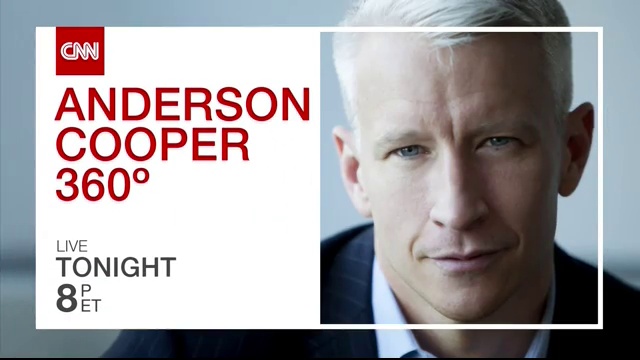}
  \includegraphics[width=0.32\columnwidth]{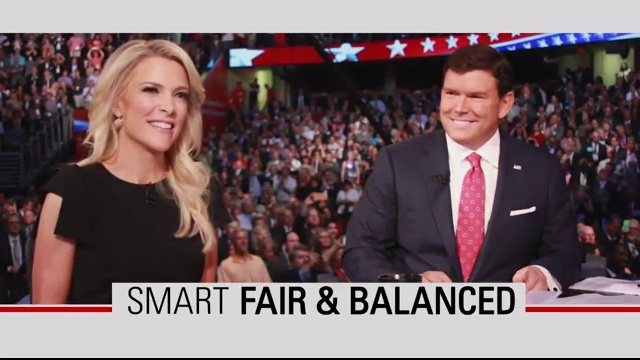}
  \includegraphics[width=0.32\columnwidth]{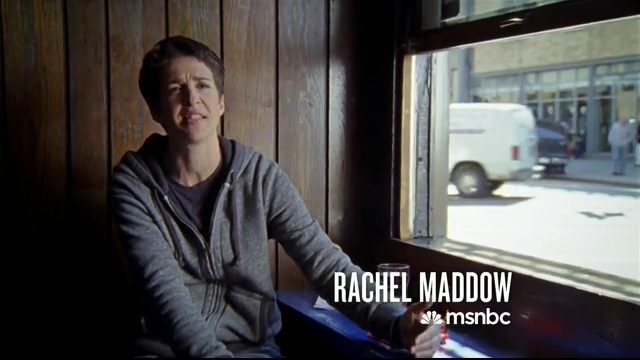}
  \caption{Example frames where news presenters (Anderson Cooper; Megyn Kelly, Bret Baier; Rachel Maddow) appear in still images and non-live video.}
  \label{fig:hosts_not_live}
\end{figure}

Note that our methodology assumes that the video was aired the same day that it
was recorded and does not account for old clips or still images
(\autoref{fig:hosts_not_live}).

\FloatBarrier
\subsection{Hair color for news presenters}

\begin{figure}[!tbp]
  \centering
  \begin{subfigure}[b]{\columnwidth}
    \centering
    \includegraphics[width=\columnwidth]{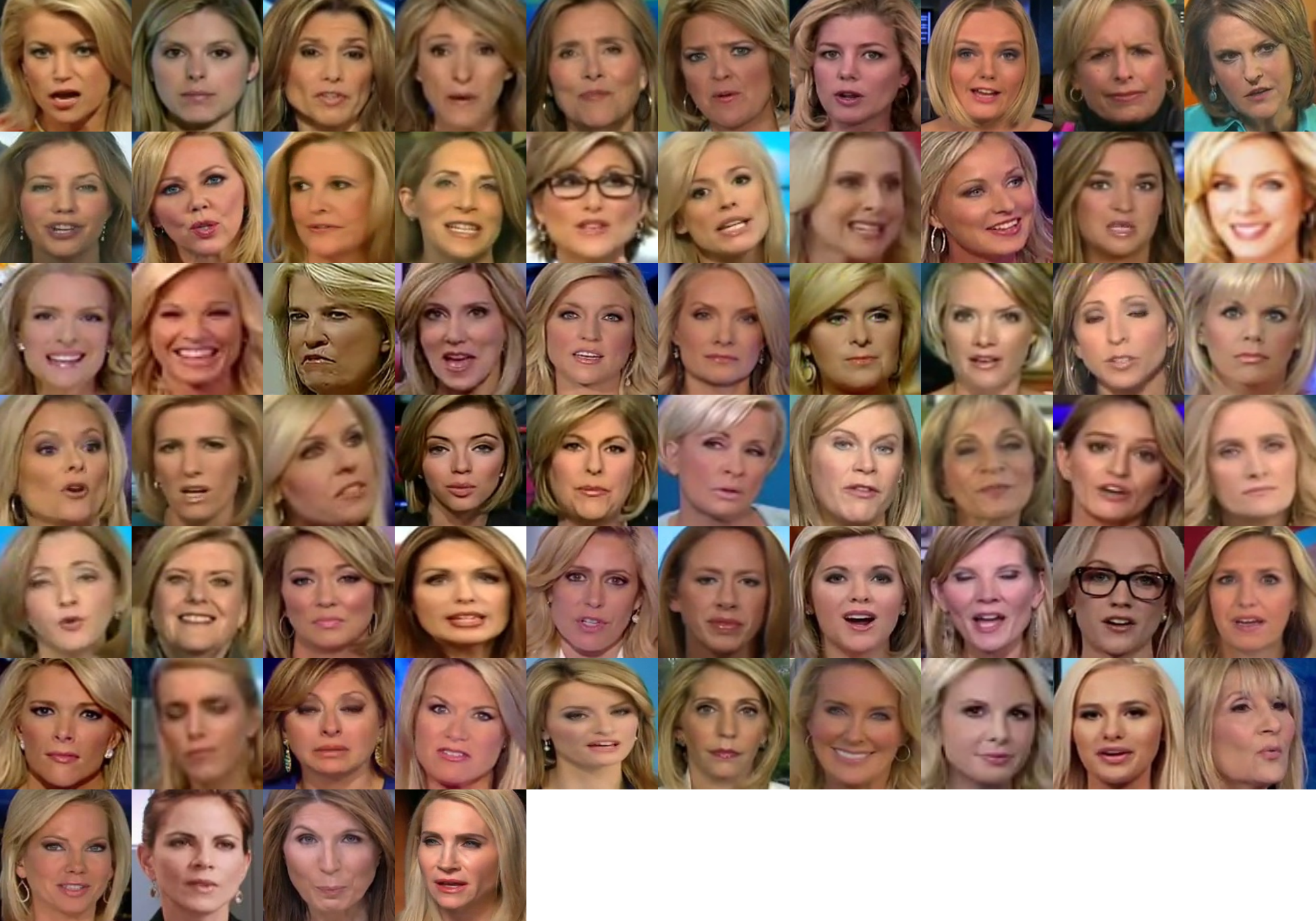}
    \caption{Blonde}
    \vspace{0.5\baselineskip}
  \end{subfigure}
  \begin{subfigure}[b]{\columnwidth}
    \centering
    \includegraphics[width=\columnwidth]{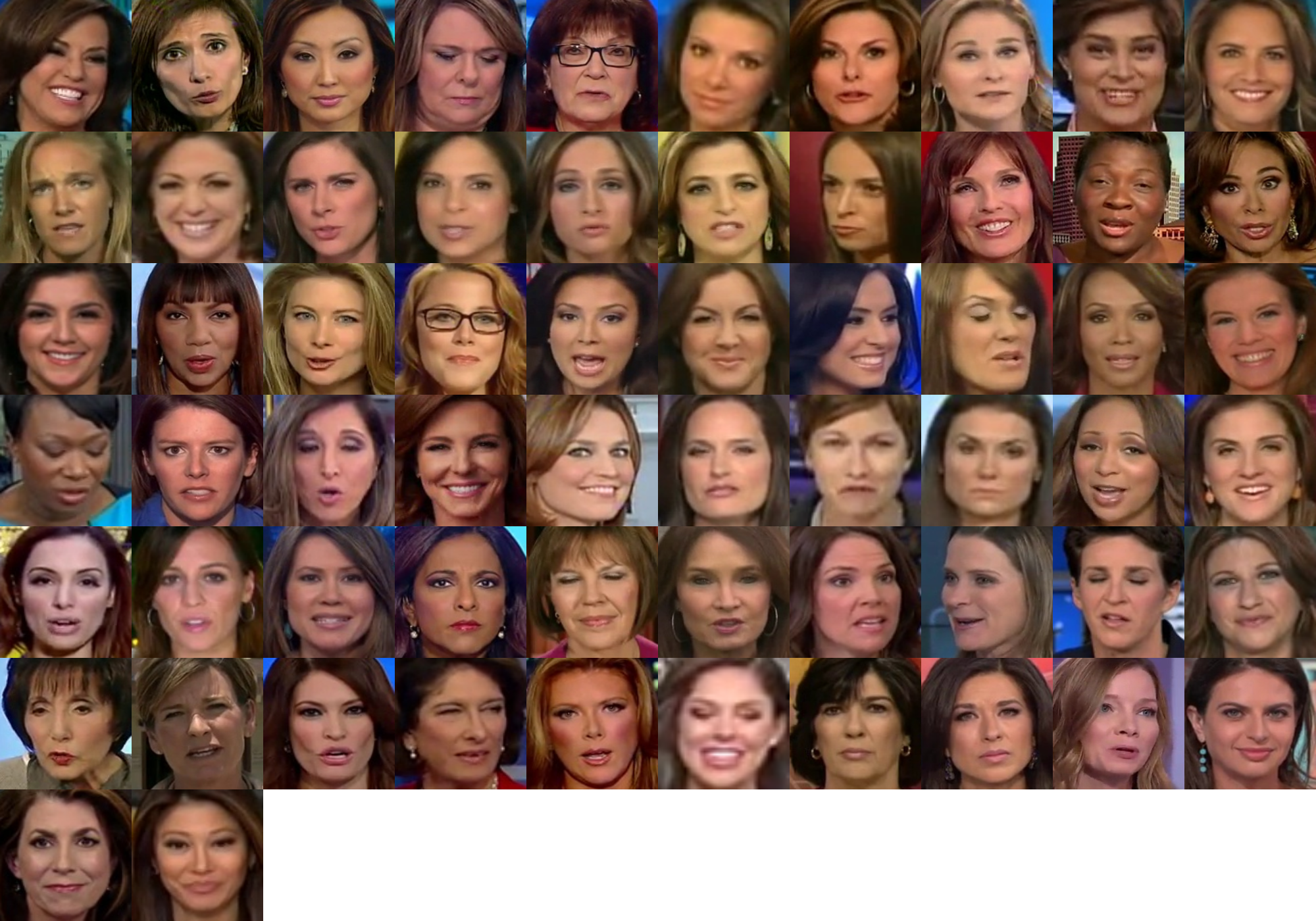}
    \caption{Brown}
    \vspace{0.5\baselineskip}
  \end{subfigure}
  \begin{subfigure}[b]{\columnwidth}
    \centering
    \includegraphics[width=\columnwidth]{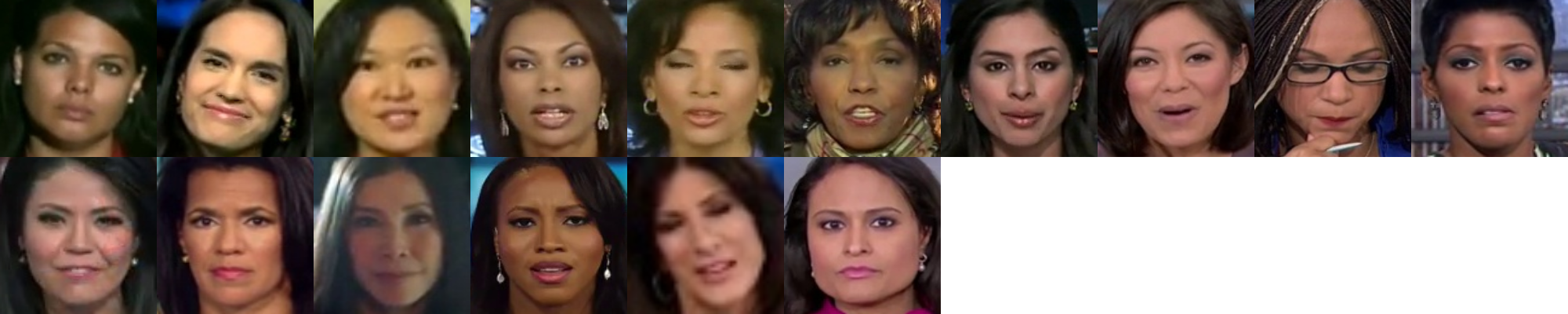}
    \caption{Black}
    \vspace{0.5\baselineskip}
  \end{subfigure}
  \begin{subfigure}[b]{\columnwidth}
    \centering
    \includegraphics[width=\columnwidth]{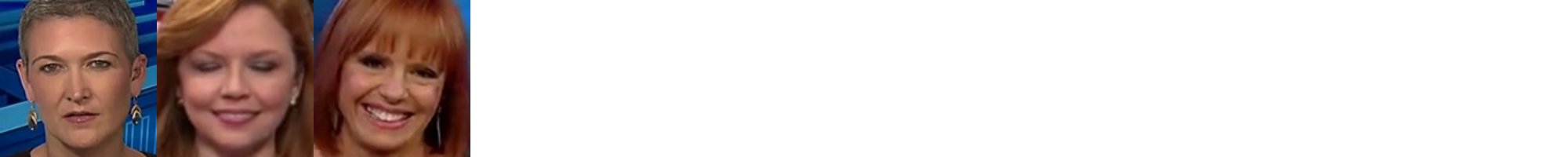}
    \caption{Other}
  \end{subfigure}
  \caption{Random image of each female-presenting news presenter, grouped by hair color label.}
  \label{fig:haircolor_examples}
\end{figure}

Two of the authors independently labeled the visible hair color for each male
and female news presenter in 25 frames sampled from the data set. There were
five possible labels (blond, brown, black, red, white/gray, and bald). For each
news presenter, we calculated the majority label according to each rater. The
inter-rater agreement for the majority label for female news presenters was
92.4\%. In these cases, the majority label was used in the analysis as the hair
color label. The two raters reviewed and agreed upon a hair color label for the
11 female news presenters where their majority labels did not match.
\autoref{fig:haircolor_examples} shows example faces from each hair color group
for the female news presenters that we analyzed.

For male presenters, the data was not analyzed because there was much lower
inter-rater agreement (75\%). One major cause of inter-rater disagreement was
confusion over when to apply the bald and white/gray hair labels. There was only
one white-haired female presenter in the data set, and no bald female
presenters, contributing to lower disagreement.

\FloatBarrier
\subsection{Images/video of Trayvon Martin and George Zimmerman}

We use our own identity labels for Trayvon Martin and George Zimmerman because
both individuals are rare overall in the data set and they are not reliably
identified by Amazon's Celebrity Recognition API~\cite{amazonrekognition}.

\begin{figure}[!tbp]
  \centering
  \begin{subfigure}[b]{0.48\columnwidth}
    \centering
    \includegraphics[width=0.3\columnwidth]{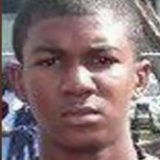}
    \includegraphics[width=0.3\columnwidth]{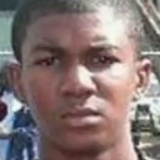}
    \includegraphics[width=0.3\columnwidth]{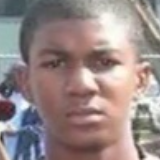}
    \caption{Image 1}
    \vspace{0.5\baselineskip}
  \end{subfigure}
  \begin{subfigure}[b]{0.48\columnwidth}
    \centering
    \includegraphics[width=0.3\columnwidth]{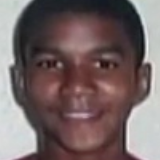}
    \includegraphics[width=0.3\columnwidth]{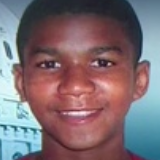}
    \includegraphics[width=0.3\columnwidth]{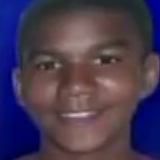}
    \caption{Image 2}
    \vspace{0.5\baselineskip}
  \end{subfigure}
  \begin{subfigure}[b]{0.48\columnwidth}
    \centering
    \includegraphics[width=0.3\columnwidth]{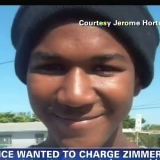}
    \includegraphics[width=0.3\columnwidth]{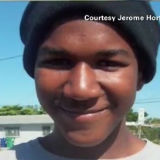}
    \includegraphics[width=0.3\columnwidth]{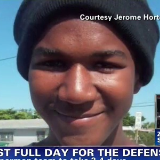}
    \caption{Image 3}
    \vspace{0.5\baselineskip}
  \end{subfigure}
  \begin{subfigure}[b]{0.48\columnwidth}
    \centering
    \includegraphics[width=0.3\columnwidth]{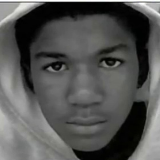}
    \includegraphics[width=0.3\columnwidth]{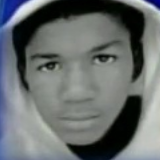}
    \includegraphics[width=0.3\columnwidth]{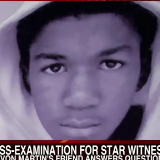}
    \caption{Image 4}
    \vspace{0.5\baselineskip}
  \end{subfigure}
  \caption{Examples of the top four images of Trayvon Martin. Images can
    have different backgrounds, color tone, sharpness, and contrast as a
    result of editing while the source image remains the same.}
  \label{fig:martin_clusters}
\end{figure}

First, we separate out faces by their source image (before any editing). In the
case of George Zimmerman, who is alive, we make a best effort to group faces
from the same source event or setting (e.g., court appearances, interview). Note
that the same image can be edited differently, have text overlays, and differ in
aspects such as tonality and background (see \autoref{fig:martin_clusters} for
examples).

\begin{figure}[!tbp]
  \centering
  \begin{subfigure}[b]{0.48\columnwidth}
    \centering
    \includegraphics[width=0.3\columnwidth]{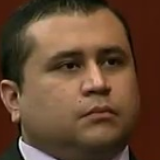}
    \includegraphics[width=0.3\columnwidth]{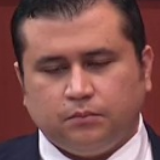}
    \includegraphics[width=0.3\columnwidth]{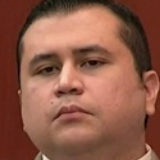}
    \caption{Video 1}
    \vspace{0.5\baselineskip}
  \end{subfigure}
  \begin{subfigure}[b]{0.48\columnwidth}
    \centering
    \includegraphics[width=0.3\columnwidth]{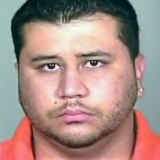}
    \includegraphics[width=0.3\columnwidth]{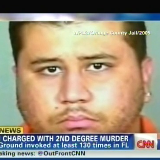}
    \includegraphics[width=0.3\columnwidth]{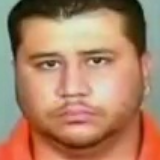}
    \caption{Image 2}
    \vspace{0.5\baselineskip}
  \end{subfigure}
  \begin{subfigure}[b]{0.48\columnwidth}
    \centering
    \includegraphics[width=0.3\columnwidth]{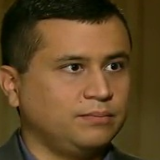}
    \includegraphics[width=0.3\columnwidth]{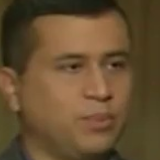}
    \includegraphics[width=0.3\columnwidth]{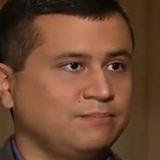}
    \caption{Video 3}
    \vspace{0.5\baselineskip}
  \end{subfigure}
  \begin{subfigure}[b]{0.48\columnwidth}
    \centering
    \includegraphics[width=0.3\columnwidth]{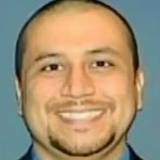}
    \includegraphics[width=0.3\columnwidth]{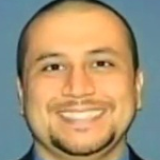}
    \includegraphics[width=0.3\columnwidth]{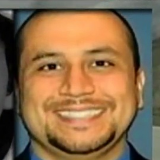}
    \caption{Image 4}
    \vspace{0.5\baselineskip}
  \end{subfigure}
  \caption{Examples of the top four image and video clusters for George Zimmerman.}
  \label{fig:zimmerman_clusters}
\end{figure}

For each individual, we use the FaceNet~\cite{facenet} descriptors (described
in~\ref{sub:face_detection}) and perform a clustering (in the embedding space)
of the faces that we previously identified as the individual. We cluster with a
human-in-the-loop, by constructing a 1-NN classifier (i.e., exact nearest
neighbor). We select faces which correspond to unique source images, partition
the faces, and then visually examine the resulting clusters. Examining the
clusters can reveal new source images or misclassified images; the human can
create new labels, fix existing labels, and repeat the process. We repeat the
process until the clusters are clean (e.g., over 90\% precise). We find that
using a 1-NN classifier is sufficient and that only a small number of manual
labels are needed (fewer than 200) to obtain good precision and recall in the
clusters (\autoref{tab:cluster_accuracy}). \autoref{fig:martin_clusters} and
\autoref{fig:zimmerman_clusters} show examples from the top four clusters for
Trayvon Martin and George Zimmerman, respectively.

\begin{table}[!tbp]
  \centering
  \begin{tabular}{lrrrr}
    \textbf{Trayvon Martin} &
    \includegraphics[width=0.10\columnwidth]{figures/portrayal/faces/martin1_1} &
    \includegraphics[width=0.10\columnwidth]{figures/portrayal/faces/martin2_1} &
    \includegraphics[width=0.10\columnwidth]{figures/portrayal/faces/martin3_1} &
    \includegraphics[width=0.10\columnwidth]{figures/portrayal/faces/martin4_1} \\
    \hline
    Precision (500 samples) & 0.996 & 0.978 & 0.988 & 0.986 \\
    Recall (500 samples) & 1.000 & 1.000 & 1.000 & 0.994 \\
    \\
    \textbf{George Zimmerman} &
    \includegraphics[width=0.10\columnwidth]{figures/portrayal/faces/zimmerman_all_1_1} &
    \includegraphics[width=0.10\columnwidth]{figures/portrayal/faces/zimmerman_all_2_1} &
    \includegraphics[width=0.10\columnwidth]{figures/portrayal/faces/zimmerman_all_3_1} &
    \includegraphics[width=0.10\columnwidth]{figures/portrayal/faces/zimmerman_all_4_1} \\
    \hline
    Contains video? & yes & no & yes & no \\
    Precision (500 samples) & 0.970 & 0.996 & 0.948 & 0.990 \\
    Recall (500 samples) & 0.941 & 1.000 & 1.000 & 1.000 \\
  \end{tabular}
  \caption{
    Estimated precision and recall for the top clusters for Trayvon Martin and
    George Zimmerman. For each cluster (say X), we estimate precision by
    sampling randomly in X and counting false positives. To estimate the number
    of false negatives (for recall) we sample faces randomly from all other
    clusters and count the number of faces that belong in cluster X, but were
    wrongly assigned. The precision estimate is used to estimate the number of
    true positives.
  }
  \label{tab:cluster_accuracy}
\end{table}

\subsection{Counting foreign country names}

To identify the set of most frequently mentioned countries, we constructed a
list of country and territory names from~\cite{countrylist}, which includes all
countries and territories with ISO-3166-1 country codes. We manually augment the
list with country name aliases; for example, the \ngram{Holy See} and
\ngram{Vatican} are aliases of one another and either term is counted as Vatican
City. A few countries such as Mexico and Georgia are substrings of U.S. state
names, leading to over-counting in the results. To address this issue, we
exclude occurrences of \ngram{Mexico} that are preceded by \ngram{New} and we
omit \ngram{Georgia} entirely. (Mentions of Georgia in U.S. cable TV news
overwhelmingly refer to the U.S. state and not the country.)

\subsection{Counting terrorism, mass shooting, and plane crash N-grams}

To measure how long the media continues to cover events after they take place, we
counted the number of times words related to terrorism, mass shootings, and
plane crashes appear following an event. \autoref{tab:major_events} and
\autoref{tab:plane_crashes} show the events that were included in the analysis.
For terrorism, we count instances of \ngram{terror(ism,ist)}, \ngram{attack},
\ngram{shooting}, \ngram{stabbing}, and \ngram{bombing}, which refer to the
attack itself; for mass shootings, the list is \ngram{shoot(ing,er)}, which
refers to the shooting or the mass shooter (searching more restrictively for
instances of \ngram{mass shoot(er,ing)} yields a similar result, but sometimes
mass is omited in the news coverage); and for plane crashes the list is
\ngram{(air)plane} or \ngram{airliner} followed by \ngram{crash} or
\ngram{missing}. Because the keywords to measure news coverage are different
between each category of event, the raw counts are not directly comparable
across categories.

\begin{table}[tp]
  \centering
  \begin{tabular}{llr}
    \textbf{Date} & \textbf{Event} & \textbf{Victims} \\
    \multicolumn{3}{l}{\textbf{Terrorist attacks (U.S.)}} \\
    \hline
    4/15/2013 & Boston Marathon bombing & 286 \\
    12/2/2015 & San Bernardino shooting & 30 \\
    6/12/2016 & Pulse Nightclub shooting & 103 \\
    9/17/2016	& 2016 New York and New Jersey & 35 \\
              & bombings \\
    8/12/2017 & Charlottesville car attack & 29 \\
    10/31/2017 & 2017 New York City truck attack & 20 \\
    8/3/2019 & El Paso shooting & 46 \\
    \\
    \multicolumn{3}{l}{\textbf{Terrorist attacks (Europe)}} \\
    \hline
    4/11/2011 & Minsk Metro bombing & 219 \\
    7/22/2011 & Norway attacks & 396 \\
    7/17/2014 & Malaysia Airlines flight 17 shootdown & 298 \\
    1/7/2015 & January 2015 Île-de-France attacks & 42 \\
    11/13/2015 & November 2015 Paris attacks & 551 \\
    3/22/2016 & Brussels bombings & 375 \\
    7/14/2016 & Nice truck attack & 521 \\
    5/22/2017 & Berlin Christmas market attack & 68\\
    6/3/2017 & Manchester Arena bombing & 273 \\
    8/17/2017 & 2017 London Bridge attack & 59\\
    2/19/2020 & 2017 Barcelona attacks & 176 \\
    \\
    \multicolumn{3}{l}{\textbf{Mass shootings}} \\
    \hline
    1/8/2011 & Tucson, Arizona & 21 \\
    7/20/2012 & Aurora, Colorado & 82 \\
    12/14/2012 & Newtown, Connecticut & 30 \\
    9/16/2013 & Washington D.C. & 21 \\
    5/23/2014 & Isla Vista, California & 20 \\
    5/17/2015 & Waco, Texas & 27 \\
    12/2/2015 & San Bernardino, California & 38 \\
    6/12/2016 & Orlando, Florida & 103 \\
    7/1/2017 & Little Rock, Arkansas & 28 \\
    10/1/2017 & Las Vegas, Nevada & 481 \\
    11/5/2017 & Sutherland Springs, Texas & 47 \\
    2/14/2018 & Parkland, Florida & 34 \\
    6/17/2018 & Trenton, New Jersey & 23 \\
    5/18/2018 & Santa Fe, Texas & 24 \\
    11/7/2018 & Thousand Oaks, California & 25 \\
    8/3/2019 & El Paso, Texas & 46 \\
    8/4/2019 & Dayton, Ohio & 37 \\
    8/31/2019 & Midland\-Odessa, Texas & 33 \\
  \end{tabular}
  \caption{Major events included in the list of terrorist attacks and mass shootings.}
  \label{tab:major_events}
\end{table}

\begin{table}[tp]
  \centering
  \begin{tabular}{llr}
    \textbf{Date} & \textbf{Plane crashes} & \textbf{Deaths} \\
    \hline
    1/25/2010 & Ethiopian Airlines Flight 409 & 90 \\
    5/12/2010 & Afriqiyah Airways Flight 771 & 103 \\
    5/22/2010 & Air India Express Flight 812 & 158 \\
    7/28/2010 & Airblue Flight 202 & 152 \\
    11/4/2010 & Aero Caribbean Flight 883 & 68 \\
    1/9/2011 & Iran Air Flight 277 & 77 \\
    7/8/2011 & Hewa Bora Airways Flight 952 & 74 \\
    4/20/2012 & Bhoja Air Flight 213 & 127 \\
    6/3/2012 & Dana Air Flight 992 & 159 \\
    11/17/2013 & Tatarstan Airlines Flight 363 & 50 \\
    3/8/2014 & Malaysia Airlines Flight 370 & 239 \\
    7/17/2014 & Malaysia Airlines Flight 17 & 298 \\
    7/24/2014 & Air Algérie Flight 5017 & 116 \\
    12/28/2014 & Indonesia AirAsia Flight 8501 & 162 \\
    3/24/2015 & Germanwings Flight 9525 & 150 \\
    8/16/2015 & Trigana Air Flight 267 & 54 \\
    3/19/2016 & Flydubai Flight 981 & 62 \\
    5/19/2016 & EgyptAir Flight 804 & 66 \\
    11/28/2016 & LaMia Airlines Flight 2933 & 71 \\
    2/11/2018 & Saratov Airlines Flight 703 & 71 \\
    2/18/2018 & Iran Aseman Airlines Flight 3704 & 66 \\
    3/12/2018 & US-Bangla Airlines Flight 211 & 51 \\
    5/18/2018 & Cubana de Aviación Flight 972 & 112 \\
    10/29/2018 & Lion Air Flight 610 & 189 \\
    3/10/2019 & Ethiopian Airlines Flight 302 & 157 \\
  \end{tabular}
  \caption{Plane crashes included in the analysis. This list includes all of the commercial airline crashes from 2010 to 2019 involving at least 50 fatalities.}
  \label{tab:plane_crashes}
\end{table}

\FloatBarrier
\subsection{Counting illegal and undocumented immigration N-grams}

We count the number of times that N-grams related to ``illegal'' and
``undocumented'' immigration appear in the captions to measure the prevalence of
both terms in discussion around immigration. The N-grams used to measure uses of
``illegal'' are \ngram{illegal immigrant(s)}, \ngram{illegal immigration},
\ngram{illegals}, and \ngram{illegal alien(s)}. For ``undocumented'', the
N-grams are \ngram{undocumented immigrant(s)}, \ngram{undocumented immigration},
and \ngram{undocumented alien(s)}.

\subsection{Counting usage of the president honorific in reference to Trump and Obama}
\label{sub:honorifics}

We measure the number of times the ``president'' honorific is used when
addressing each president. This requires classifying occurrences of the word
\ngram{Trump} (and also \ngram{Obama}) in captions as having the ``president''
honorific, not having the honorific (e.g., \ngram{Donald Trump} or just
\ngram{Trump}), or not referring to his person (e.g., Trump University).

For Donald Trump, we only count exact matches of \ngram{President Trump} or
\ngram{President Donald Trump} as uses of ``president''. To count occurrences of
without the honorific, we exclude occurrences preceded by \ngram{president} and
instances followed by \ngram{administration}, \ngram{campaign},
\ngram{university}, and \ngram{care}, which are used in compound nouns with
\ngram{Trump}. We also exclude occurrences preceded by \ngram{the} (e.g., to
filter out other compound nouns of the form \ngram{the Trump ...}); note that
this also removes \ngram{the Trump presidency}, which is not referring to his
person, but his presidency. Finally, we exclude Donald Trump's immediate family:
\ngram{Melania}, \ngram{Ivanka}, \ngram{Eric}, \ngram{Barron}, and [Donald
Trump]~\ngram{Jr}. These exclusions of nouns related to Trump (but not directed
at his person) were selected by visual examination of the top 100 bigrams
containing \ngram{Trump}.

The methodology for counting references to Barack Obama is identical, except
that the excluded family members are \ngram{Michelle}, \ngram{Malia}, and
\ngram{Sasha}.

\subsection{Measuring visual association between words and male/female-presenting screen time}
\label{sub:gender_text}

We compute the conditional probabilities of any male- or any female-presenting
face being on screen when a word appears in the text.

The majority of the words in the data set (including rare words, but also
misspellings) occur very infrequently -- 95.6\% of unique tokens appear fewer
than 100 times in the data set. Because there are few face detection events
corresponding to these words, their conditional probability has high variance,
often taking on extreme values. In order to remove these words and to make the
computation practical, we considered only words that appear at least 100 times
in the captions.

From the remaining tokens, we filter out NLTK English stop words~\cite{nltk}
and restrict our analysis to the most common words in the data set, considering
only the top 10\% of remaining words (words that occur over 13,462 times).

We then rank the words according to the difference in conditional probability of
female-presenting and male-presenting faces given the word appearing in the
caption. The top and bottom words in this list are the most strongly associated
with the two presented genders. We report the top 35 words for each presented
gender, manually filtering out words in these lists that are human names (e.g.,
\ngram{Alisyn} is associated with female-presenting screen time because Alisyn
Camerota is a presenter on CNN) or news program names (which associate to the
genders of hosts).

The top female-associated word, \ngram{futures} is similar to other
highly-ranked words in the list (\ngram{NASDAQ, stocks}, but is also part of the
name of a female-hosted TV program ({\it Sunday Morning Futures}). 14.6\%
percent of \ngram{futures} mentions are part of the 3-gram \ngram{sunday morning
futures} The word with the 14th-highest conditional probability \ngram{newsroom}
is also both a common news-related word and part of a news program name
({\it CNN Newsroom}).

\subsection{Computing unique words for individuals}

To determine which individuals and words have strong visual/textual
associations, we compute the amount of time each individual was on screen while
each word is said. This is used to calculate the conditional probability that a
person is on screen given the word being said. To filter out rare words, we only
consider words with at least 100 occurrences across the decade. The words with
conditional probabilities exceeding 50\% for any individual are given in Table. 1
in the paper.

\subsection{Measuring visual association between news presenters and the president honorific}

We extended the president honorific analysis (methodology in
\ref{sub:honorifics}) to when various news presenters are on screen. The N-grams
that are counted remain the same as in~\ref{sub:honorifics}.  We start with the
list of news presenters described in~\ref{sub:presenters}, but we only show news
presenters with at least 100 total references to Trump and 100 total references
to Obama to ensure that there is sufficient data for a comparison. This is to
account for news presenter who retired before Trump became president or started
after Obama stepped down.

\subsection{Measuring visual association between Clinton and the word email}

The Hillary Clinton email scandal and subsequent FBI investigation was a highly
polarizing issue in the 2016 presidential election. To measure the degree to
which Clinton is visually associated with the issue, represented by
the word ``email'', we counted the number of times ``email(s)'' was said, and the
number of times it was said while Clinton are on screen.

We count occurrences of \ngram{e mail(s)}, \ngram{email(s)}, and
\ngram{electronic mail} as instances of email being said in the captions. There
are 122K utterances of email in the captions between 2015 and 2017, while
Hillary Clinton has 738 hours of screen time in the same time period. Clinton's
face is on screen during 14,019 of those utterances.

\FloatBarrier
\subsection{Detecting interviews}
\label{sec:interviews}

Our algorithm for interviews in TV News searches for interviews between a news
presenter (the host) and a named guest X. We search for segments where the guest and
the host appear together, surrounded by the guest appearing alone or the host
appearing alone. Combining these segments captures an alternating pattern where
a host appears, guest appears, ... that is indicative of an interview. The pseudocode
for this algorithm is shown in Rekall~\cite{rekall} in
\autoref{fig:code_interview}.

\begin{figure}[!tbp]
  \centering
  \begin{subfigure}[b]{\columnwidth}
    \centering
    \includegraphics[width=0.24\columnwidth]{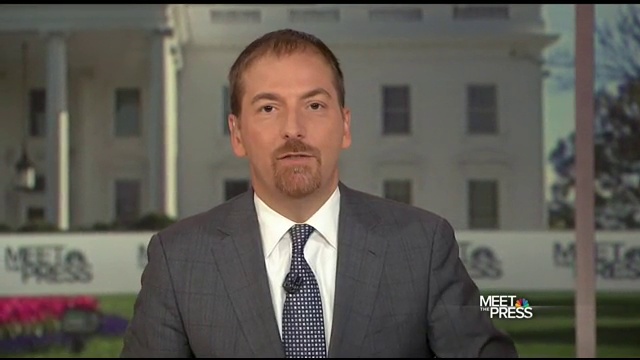}
    \includegraphics[width=0.24\columnwidth]{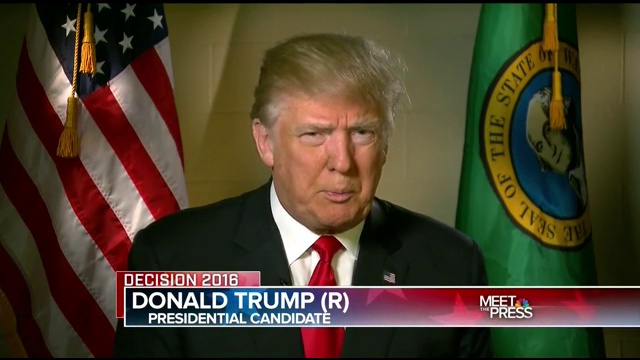}
    \includegraphics[width=0.24\columnwidth]{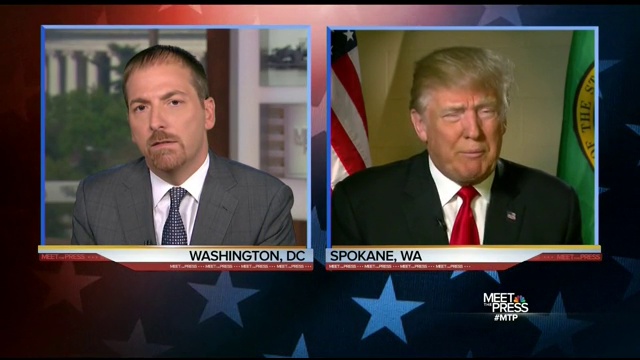}
    \includegraphics[width=0.24\columnwidth]{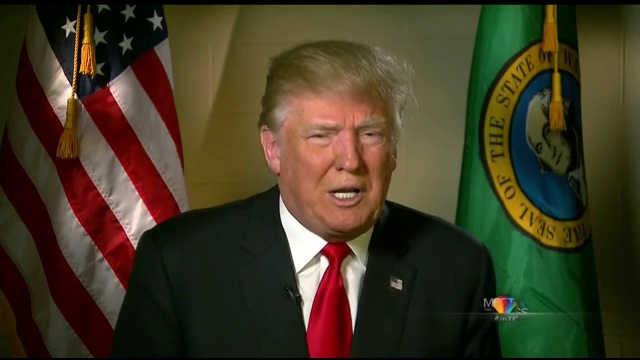}
    \caption{A real interview.}
    \vspace{0.5\baselineskip}
  \end{subfigure}
  \begin{subfigure}[b]{\columnwidth}
    \centering
    \includegraphics[width=0.24\columnwidth]{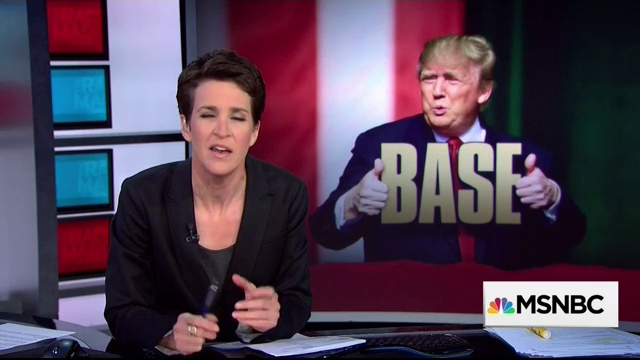}
    \includegraphics[width=0.24\columnwidth]{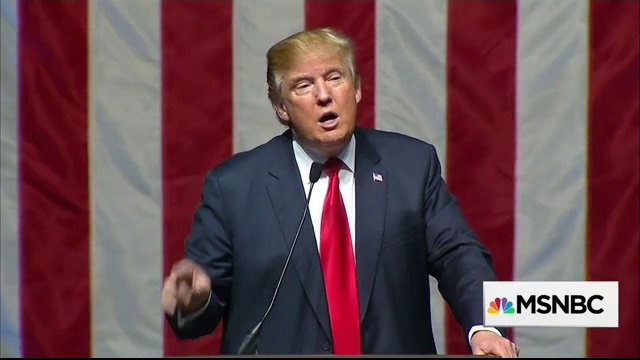}
    \includegraphics[width=0.24\columnwidth]{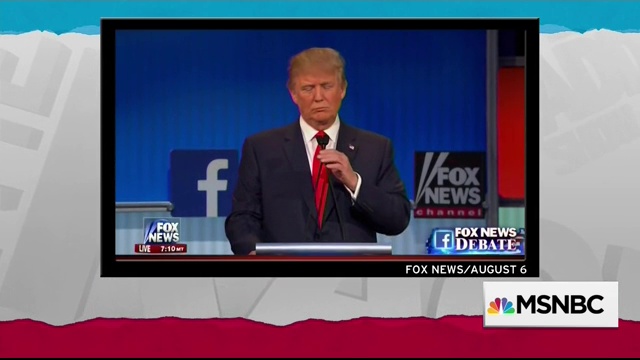}
    \includegraphics[width=0.24\columnwidth]{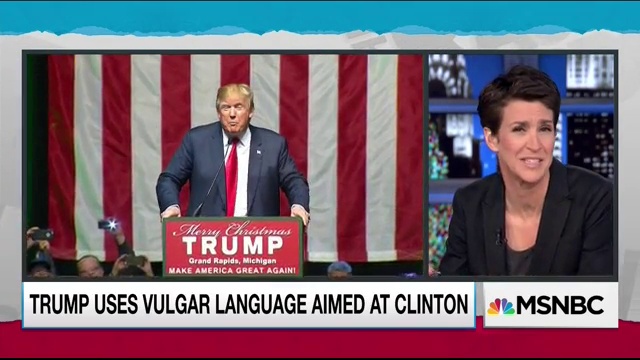}
    \caption{Not an interview.}
    \vspace{0.5\baselineskip}
  \end{subfigure}
  \caption{Example frames from a real and incorrectly detected interview. Note that both follow a pattern of a host and guest being on screen, together and alone. The incorrectly detected interview contains videos and graphics of Donald Trump in lieu of his live person. As the presidents and leading candidates, Trump, Clinton, and Obama are discussed at length by hosts in visual contexts that appear similar to interviews.}
  \label{fig:interview_examples}
\end{figure}

We applied this interview detection algorithm on 44 people across our whole
data set. These individuals are listed in~\autoref{tab:interview_candidates}.

We exclude Barack Obama, Donald Trump, and Hillary Clinton due to those
individuals appearing too often in video clips and still images. Their
appearances along with hosts are often misclassified as interviews. For example,
Donald Trump may be shown in a still image or giving a speech while the news
content cuts back and forth to a host providing
commentary (\autoref{fig:interview_examples}). Events such as town-hall gatherings
are sometimes also confused as interviews. As the leading candidates and
presidents, Trump, Clinton, and Obama appear the most often in these contexts.

We validated our interview detection algorithm by annotating 100 cable TV news
videos which contain interviews for three interviewees: Bernie Sanders,
Kellyanne Conway, John McCain. \autoref{tab:interview_precision_recall} shows
the estimated precision and recall numbers for the three interviewees, as well
as the total amount of interview screen time in ground truth for each
interviewee.

\begin{figure}[htp]
\begin{lstlisting}
# Interviews between a host and a named guest
faces = rekall.ingest(database.table("faces"), 3D)

# Select all faces (3s segments) identified as the
# guest and the faces of all hosts
guest_faces = faces.filter(
  face: face.name = guest_name)
host_faces = faces.filter(
  face: face.is_host)

# Coalesce adjacent segments since individuals are
# often on screen for longer than the 3s sample rate
guest_segs = guest_faces.coalesce(
  predicate = time_gap < 30s,
  merge = time_span)
host_segs = host_faces.coalesce(
  predicate = time_gap < 30s,
  merge = time_span)

# Find segments when a host and the guest are on
# screen at the same time
guest_and_host_segs = guest_segs.join(
  host_segs,
  predicate = time_overlaps,
  merge = time_intersection)

# Find segments when the guest is on screen without
# the host
guest_alone_segs = guest_segs.minus(
  guest_and_host_segs)

# Merge segments when the guest is on screen alone
# with the segments when both the host and guest are
# on screen and consider these to be segments of
# an interview
interview_segs = guest_and_host_segs.join(
  guest_alone_segs,
  predicate = before or after,
  merge = time_span)

# Merge the detected interview segments and return
# the ones that exceed a minimum interview duration
interviews = interview_segs
  .coalesce()
  .filter(interval:
     interval["t2"] - interval["t1"] >= 240s)
\end{lstlisting}
\caption{
Rekall\,\cite{rekall} query to retrieve interviews between a host and a named guest (e.g., Bernie Sanders).
}
\label{fig:code_interview}
\end{figure}

\begin{table}[htp]
  \centering
  \begin{tabular}{lr}
    \textbf{Interviewee} & \textbf{Hours} \\
    \hline
    John McCain & 124.4 \\
    Bernie Sanders & 107.8 \\
    Rand Paul & 98.0 \\
    Lindsey Graham & 93.3 \\
    Rick Santorum & 91.9 \\
    Marco Rubio & 87.9 \\
    Kellyanne Conway & 77.7 \\
    Sarah Palin & 72.0 \\
    Paul Ryan & 67.5 \\
    John Kasich & 63.5 \\
    Ted Cruz & 61.5 \\
    Chris Christie & 61.5 \\
    Mitt Romney & 58.9 \\
    Ben Carson & 49.1 \\
    Elizabeth Warren & 35.4 \\
    Mitch McConnell & 34.7 \\
    Carly Fiorina & 33.7 \\
    Cory Booker & 31.3 \\
    Kevin McCarthy & 31.0 \\
    Tim Kaine & 29.4 \\
    Chuck Schumer & 28.9 \\
    Nancy Pelosi & 28.9 \\
    Amy Klobuchar & 28.5 \\
    Jeb Bush & 26.8 \\
    Dick Durbin & 25.8 \\
    John Boehner & 24.6 \\
    Joe Biden & 24.2 \\
    Bill Clinton & 22.0 \\
    Bill De Blasio & 19.6 \\
    George W. Bush & 19.2 \\
    Steve Scalise & 18.2 \\
    Bobby Jindal & 17.3 \\
    Orrin Hatch & 15.1 \\
    Martin O'Malley & 14.6 \\
    Kamala Harris & 12.9 \\
    John Cornyn & 10.3 \\
    Tulsi Gabbard & 9.6 \\
    Harry Reid & 7.6 \\
    Pete Buttigieg & 7.5 \\
    Jim Webb & 6.1 \\
    Beto O'Rourke & 5.3 \\
    Lincoln Chafee & 4.4 \\
    Michelle Obama & 2.3 \\
    Jim Gilmore & 1.6 \\
    \hline
    Newt Gingrich & 185.3 \\
    Mike Huckabee & 95.8 \\
  \end{tabular}
  \caption{Detected interview time for prominent U.S. political figures. Newt Gingrich and Mike Huckabee
  are listed separately because they are both hosts (news presenters) and politicians.}
  \label{tab:interview_candidates}
\end{table}

\begin{table}[!tbp]
  \centering
	\begin{tabular}{lrrr}
		\textbf{Interviewee} & \textbf{Hours} & \textbf{Precision} & \textbf{Recall} \\ \hline
		Bernie Sanders    & 3.5 & 91.7\% & 97.5\% \\
		Kellyanne Conway  & 2.2 & 91.8\% & 89.1\% \\
		John McCain       & 0.9 & 86.0\% & 99.5\% \\
	\end{tabular}
	\caption{Precision and recall numbers for the interview detector across 100 hand-annotated videos as well as the total amount of interview screen time in ground truth for each interviewee.}
	\label{tab:interview_precision_recall}
\end{table}

\FloatBarrier
\newpage
\begin{figure*}[htbp]
  \centering
  \includegraphics[width=\textwidth]{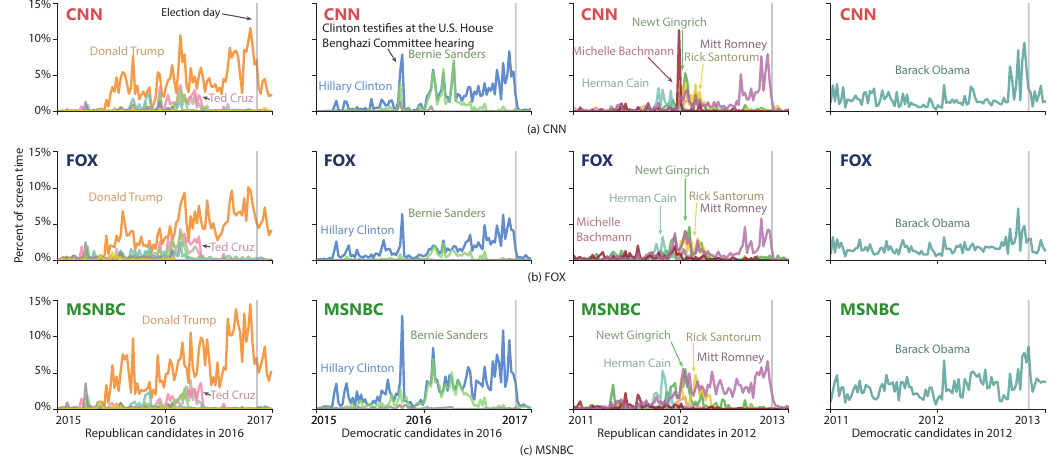}

  \caption{Donald Trump received more screen time than any other Republican
  candidate in the 2016 election season. The difference is most pronounced on
  MSNBC. Hillary Clinton and Bernie Sanders received similar amounts of screen
  time during the competitive period of the presidential primary season (January
  to May, 2016). Compared to CNN and MSNBC, FOX gave less screen time to the
  Democratic candidates in 2016. In the 2012 election season, Mitt Romney did
  not dominate screen time of the Republican candidates until much later in the
  primary season. Michelle Bachmann received a much larger peak on CNN in
  January 2012 on CNN (compared to FOX and MSNBC) before the Iowa caucuses and
  after she dropped out of the race. Finally, both Barack Obama, the incumbent
  Democratic president, and Mitt Romney received more screen time on MSNBC than
  on CNN and FOX.
  }

  \label{fig:primaries_by_channel}
\end{figure*}

\section{Additional analyses}

\subsection{Who is in the news?}

\subsubsection{How much time is there when at least one face is on screen in commercials?}
Recall from the paper that the percentage of screen time when a face is on
screen in news content has risen from by 8.6\%, from 72.9\% in 2010 to 81.5\% in
2019. This same percentage has only risen slightly in commercials in the same
timespan (38\% to 41\%), suggesting that the increase is not solely due to
improvements in video quality.

\begin{figure}[!tbp]
  \centering
  \includegraphics[width=\columnwidth]{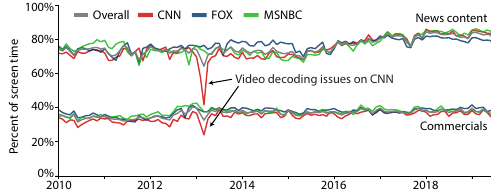}
  \caption{The percentage of time when faces are on screen has increased for news content, but has remained static in commercials since 2013.}
  \label{fig:avg_num_faces_time_by_channel}
\end{figure}

The average number of detected faces visible on screen is 1.38 in news content
and 0.49 in commercials, and these figures vary little between channels. There
is a rise in the number of detections over the decade, across all three
channels, from 1.2 in 2010 to 1.6 in 2019, with much of the increase since 2015
(\autoref{fig:avg_num_faces_time_by_channel}). By contrast, the average number
of faces on screen in commercials rises from 0.42 to 0.52, with the much of the
increase occurring before 2012.

\begin{figure}[!tbp]
  \centering
  \includegraphics[width=\columnwidth]{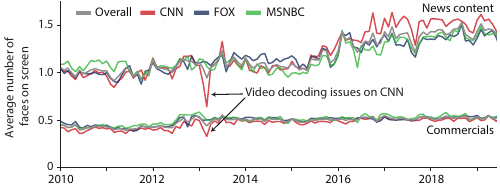}
  \caption{The average number of faces on screen has increased on all three channels.}
  \label{fig:avg_num_faces_time_by_channel}
\end{figure}

\subsubsection{What is the average size of faces?}
\begin{figure}[!tbp]
  \centering
  \includegraphics[width=\columnwidth]{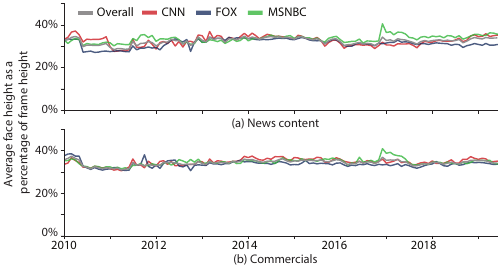}
  \caption{The average height of faces on screen has remained mostly constant in both news content and commercials, but there is some variation within the decade. The average heights of faces in news content and commercials are similar.}
  \label{fig:face_height_time_by_channel}
\end{figure}

The average size of detected faces in news content, as a proportion of
the frame height has also risen slightly from 33\% to 35\% on CNN and 33\% to
36\% on MSNBC, but has fallen from 33\% to 31\% on FOX
(\autoref{fig:face_height_time_by_channel}a). Within commercials, the change is
less than 1\% on CNN and MSNBC, but has fallen from 38\% to 34\% on FOX
(\autoref{fig:face_height_time_by_channel}b). Note that some videos have black
horizontal bars on the top and the bottom due to the video resolution not
matching the aspect ratio as an artifact of the recording (16:9 inside 4:3). We
excluded these black bars from the frame height calculation.

\subsubsection{Did the screen time given to presidential candidates vary by
channel?} There is some variation in the screen time given to candidates across
channels, but the overall patterns are similar to the aggregate patterns
described in the paper (\autoref{fig:primaries_by_channel}).

% Lifted from below
\begin{figure*}[!tbp]
  \centering
  \includegraphics[width=\textwidth]{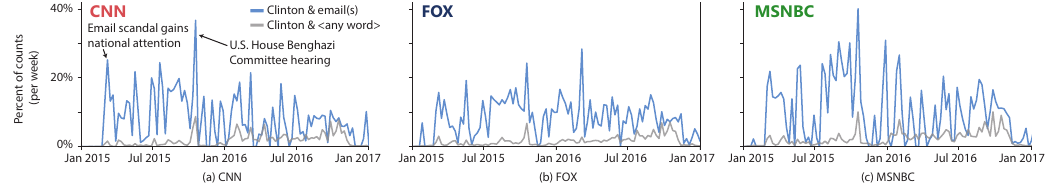}
  \caption{The visual association between Hillary Clinton's face and the word ``emails'' follows a similar trend on all three channels, far exceeding the baseline association between Clinton being on screen and any arbitrary word being said. From July to October, 2015, Clinton is shown the most on MSNBC (peaking at 40\%) when \ngram{email} is said.}
  \label{fig:emails_by_channel}
\end{figure*}

\subsubsection{Do shows presented by female-presenting news presenters give more screen time to women overall?}
An individual show's overall gender balance is skewed by the gender of its host.
For example, the show with the greatest female-presenting screen time is {\it Melissa
Harris-Perry} on MSNBC and the show with the greatest male screen time is {\it
Glenn Beck} on FOX.

We use the percentage of female-presenting news presenter screen time out of
total news presenter screen time to measure the extent to which a show is
female- or male-presented. As a measure of the gender balance for
female-presenting individuals who are not presenters (non-presenter), we compute
the percentage of female-presenting screen time for faces not identified as a
news presenter out of the time for all faces that are not identified as a
presenter. We measured the linear correlation between these two percentages to
evaluate whether shows that lean toward more female-presenting news presenter
screen time also have more screen time for female-presenting non-presenters in
general. To exclude short-lived shows and special programming, we limited the
analysis to shows with at least 100 hours of news content.

\begin{figure}[!ttbp]
  \centering
  \includegraphics[width=\columnwidth]{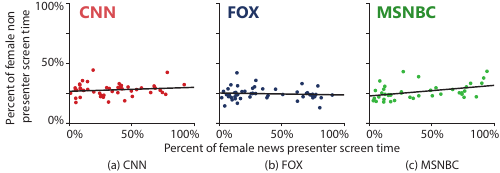}
  \caption{There is little correlation between shows that are predominantly presented by female-presenting news
  presenters and shows with the most screen time for female-presenting faces who are not news
  presenters.}
  \label{fig:female_host_show}
\end{figure}

We find no correlation on CNN ($slope=0.03, R^2=0.02$) and FOX ($slope=-0.02,
R^2=0.01$), and a weak positive correlation on MSNBC ($slope=0.09, R^2=0.19$)
(\autoref{fig:female_host_show}). This suggests that shows hosted by
female-presenting news presenters do not give proportionally more screen time to
female-presenting subjects and guests. Our result contrasts with findings by the
GMMP\,\cite{gmmp} that female journalists write disproportionately more articles
about female subjects.

\subsubsection{Which politicians get interviewed? Which presenters do
interviews?} Interviews are one of the ways that cable TV news channels bring on
experts and provide politicians with a platform to express their views. We find
interviews by looking for continuous segments of video when a presenter
(interviewer) and interviewee are on screen together and/or alternating back and
forth (details in~\autoref{sec:interviews}). Empirically, we found that this
identifies interview segments for 44 prominent American political figures that
we tested (including 17 2016 US presidential candidates). (Note: we exclude
Barack Obama, Mitt Romney, Donald Trump, and Hillary Clinton because they appear
too frequently in non-interview contexts, leading to low precision in detecting
interviews. Newt Gingrich and Mike Huckabee, who are both hosts and political
figures, are also excluded.)

In the interviews that we detected, John McCain is featured the most. Many of
the top interviewees among the individuals that we tested are Republicans. This
is due to our biased sampling toward 2016 presidential candidates and the
relatively competitive and crowded Republican primary (compared to the
Democratic primary that year). The top three interviewers are all hosts on FOX;
Greta van Susteren (former host of {\it On the Record} on FOX) is the most
prolific.

\begin{figure}[!tbp]
  \centering
  \includegraphics[width=\columnwidth]{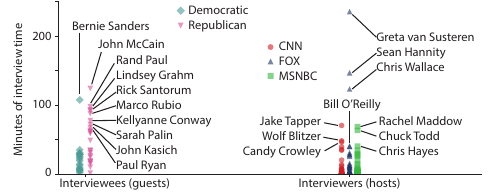}
  \caption{Interview time of the 44 politicians (interviewees) tested and hosts (interviewers). Note: Bernie Sanders is labeled Democratic due to his affiliation in the 2016 primary.}
  \label{fig:interview_time}
\end{figure}

\subsubsection{What is the visual layout of interviews?}
\begin{figure}[!tbp]
  \centering
  \includegraphics[width=\columnwidth]{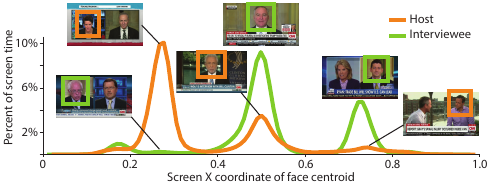}
  \caption{In interviews, the host appears overwhelmingly on the left or in the middle;
  interviewees appear in the middle or on the right.}
  \label{fig:where_is_the_head}
\end{figure}

In the majority of interviews, the host appears on the left (split-screen) or in
the middle, while the interviewee typically appears on the right (split-screen)
or in the middle (\autoref{fig:where_is_the_head}). This is in contrast to late night
talk shows, which place the host on the right.

\subsection{What is discussed?}

\subsubsection{Does any channel cover foreign countries more than the others?}
\begin{figure}[!tbp]
    \centering
    \includegraphics[width=\columnwidth]{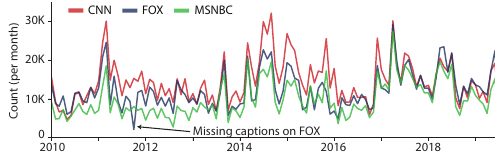}
    \caption{The number of times when foreign country names appear in the news
    oscillates. The peaks on all three channels are concurrent, but until 2017, the count of foreign country names was higher on CNN than on FOX and MSNBC.}
    \label{fig:foreign_country_count}
\end{figure}

The number of times that foreign countries appear in the text captions
oscillates over time, likely due to major events occurring abroad
(\autoref{fig:foreign_country_count}). However, all three channels follow a
similar trajectory.

\subsection{Who is on screen when a word is said?}

\subsubsection{Did different channels visually associate Hillary Clinton more
with the word \ngram{email} than others?}

\autoref{fig:emails_by_channel} shows the percentage of times when \ngram{email}
is said and when Hillary Clinton is on screen.

\FloatBarrier
\newpage

\bibliographystyle{ACM-Reference-Format}
\bibliography{tvnews}